\documentclass[11pt]{article}
\pdfoutput=1
\usepackage{empheq}

\usepackage{amsmath, amsfonts, amssymb}
\usepackage{comment}
\usepackage{graphicx}
\usepackage{psfrag}
\usepackage[usenames,dvipsnames,svgnames,table]{xcolor}
\usepackage{enumerate}
\usepackage{soul}
\usepackage{subfig}
\usepackage{cite}
\usepackage{a4wide}
\usepackage{tikz}

\usepackage{color}
\definecolor{dark-gray}{gray}{0.20}
\definecolor{gray}{gray}{0.30}
\definecolor{light-gray}{gray}{0.80}
\definecolor{dark-red}{rgb}{0.7,0,0}
\definecolor{dark-green}{rgb}{0.1,0.4,0}
\definecolor{dark-blue}{rgb}{0.3,0.3,0.7}
\definecolor{light-blue}{rgb}{0.8,0.8,1}

\usepackage{hyperref}
\usepackage{setspace}
\hypersetup{
    colorlinks=true,
    linkcolor=dark-blue,
    citecolor=dark-red,
    urlcolor=dark-green,
    linktoc=page
}

\numberwithin{equation}{section}

\def\be{\begin{equation}}
\def\ee{\end{equation}}
\def\bea{\begin{eqnarray}}
\def\eea{\end{eqnarray}}

\renewcommand{\L}{\mathcal{L}}
\newcommand{\R}{\mathbb{R}}
\newcommand{\Z}{\mathbb{Z}}
\newcommand{\N}{\mathcal{N}}

\DeclareMathOperator{\Tr}{Tr}
\DeclareMathOperator{\GL}{GL}
\DeclareMathOperator{\SL}{SL}
\DeclareMathOperator{\SO}{SO}
\DeclareMathOperator{\SU}{SU}
\DeclareMathOperator{\U}{U}
\renewcommand{\O}{\mathrm{O}}

\newcommand{\AdS}{\mathrm{AdS}}
\newcommand{\vol}{\mathrm{vol}}

\renewcommand{\d}[2][]{\mathrm{d}^{#1}{#2}}
\newcommand{\der}[3][]{\frac{\mathrm{d}^{#1}{#2}}{\mathrm{d}{#3}^{#1}}}

\renewcommand{\c}{\mathsf{c}}

\usepackage{letltxmacro}
\makeatletter
\let\oldr@@t\r@@t
\def\r@@t#1#2{%
\setbox0=\hbox{$\oldr@@t#1{#2\,}$}\dimen0=\ht0
\advance\dimen0-0.2\ht0
\setbox2=\hbox{\vrule height\ht0 depth -\dimen0}%
{\box0\lower0.4pt\box2}}
\LetLtxMacro{\oldsqrt}{\sqrt}
\renewcommand*{\sqrt}[2][\ ]{\oldsqrt[#1]{#2}}
\makeatother

\begin{document}

\hfill{KEK-TH-2495}\par
\smallskip
{\vskip26pt plus0.4fil minus15pt}

\begin{center}
    {\LARGE\bf A 10d construction of Euclidean axion\\ \vspace{0.35cm} wormholes in flat and AdS space}\\

    \vspace{1.5cm}
    {\large Gregory J.\ Loges$^1$, Gary Shiu$^2$, Thomas Van Riet$^3$}\\
    \vspace{1cm}

    {$^1$ \emph{
        Theory Center, IPNS, High Energy Accelerator Research Organization (KEK),\\
        1-1 Oho, Tsukuba, Ibaraki 305-0801, Japan
    }}\\ \vspace{0.15cm}
    {$^2$ \emph{
        Department of Physics, University of Wisconsin-Madison, Madison, 53706, USA
    }}\\ \vspace{0.15cm}
    {$^3$ \emph{
        Instituut voor Theoretische Fysica, K.U. Leuven,\\
        Celestijnenlaan 200D B-3001 Leuven, Belgium
    }}
    
    \vspace{2cm}
    
    {\bf Abstract}
\end{center}

{\small Euclidean wormhole geometries sourced by axions and dilatons are puzzling objects in quantum gravity. From one side of the wormhole to the other, the scalar fields traverse a few Planck lengths in field space and so corrections from the UV might potentially affect the consistency of the solution, even when the wormholes are large. Motivated by this, we carry out the first explicit 10d lifts of regular Euclidean axion wormholes. We start off with the lift of Giddings-Strominger wormholes in $\N=8$ Euclidean supergravity over a 6-torus to 10d type IIA supergravity and find the solution can be everywhere tuned into the parametrically controlled supergravity regime. Secondly, we construct explicit wormholes in AdS spaces and find them again to be under parametric control. We find the first wormhole solutions in massive type IIA on $S^3\times S^3$ and in type IIB on $T^{1,1}$. The latter has an explicit holographic dual, and similar to the earlier constructions in $\AdS_5 \times S^5/\Z_k$, the wormholes violate operator positivity since $\Tr(F\pm{\star F})^2<0$. This puzzle might arise from subtleties related to computing holographic $n$-point functions in the presence of multiple boundaries.
}

\setcounter{tocdepth}{2}

\newpage
\tableofcontents


\section{Introduction}
\label{sec:intro}

The study of Euclidean wormholes in quantum gravity, reviewed in~\cite{Hebecker:2018ofv}, has seen a revival in the recent decade. This is mainly due to our improved understanding of the path integral for low-dimensional gravity~\cite{Mertens:2022irh} and the growing interest in the Swampland program~\cite{Vafa:2005ui}. This growing body of work points to wormholes being relevant saddle points, yet an understanding within higher-dimensional gravity is more difficult and it has even been suggested that in $d>3$ wormholes are in some sense spurious~\cite{McNamara:2020uza, VanRiet:2020pcn}. In particular, it requires an understanding of Euclidean wormholes within string theory and holography~\cite{Kundu:2021nwp} and hence an embedding of wormhole solutions, preferably, into 10d supergravity. 

The most straightforward method to embed such wormholes relies on axion fields, as they rather naturally provide the required negative Euclidean energy-momentum tensor to source wormhole geometries~\cite{Giddings:1987cg}.\footnote{Other kinds of Euclidean wormholes have been studied in a holographic context in~\cite{Betzios:2019rds, Betzios:2021fnm,Marolf:2021kjc,Maldacena:2004rf}.} A possible interpretation of these Giddings-Strominger (GS) wormholes is as instantons describing the birth or absorption of baby universes~\cite{Coleman:1988cy}. Since these baby universes carry axion charge, it would lead to an apparent violation of axion charge conservation, which is expected on general grounds since axion shifts are global symmetries of which only a discrete shift is gauged. Indeed, wormholes could then be the instantons that generate Planck-suppressed potential terms for the axion~\cite{Kallosh:1995hi}. In this context, GS wormholes have been of great interest in the Swampland program since one can wonder whether they destroy very flat inflaton potentials required for large field inflation. This relates to GS wormholes being the objects predicted by the axion version of the Weak Gravity Conjecture~\cite{ArkaniHamed:2006dz, Montero:2015ofa, Brown:2015iha, Brown:2015lia, Heidenreich:2015nta, Hebecker:2016dsw, Hebecker:2017wsu,Hebecker2017, Guidetti:2022xct}. Indeed, GS wormholes in the presence of an axion can be seen as the over-extremal partners of BPS instanton-like configurations~\cite{Bergshoeff:2004fq, VanRiet:2020pcn}. Yet the ultimate fate of these wormholes depends, among other things, on their stability, which, despite various papers on the topic, was never really settled until recently in~\cite{Loges:2022nuw}. There it was shown that GS wormholes in the absence of a dilaton are perturbatively stable. The fate of wormhole stability in the presence of both axions and dilatons is work in progress.

Surprisingly, it has taken a long time to embed GS-like wormholes into string theory, and to our knowledge the first proof for the existence of regular GS wormholes in string theory appeared in~\cite{Bergshoeff:2004pg}. Their construction leads to regular solutions in four flat Euclidean dimensions and applies to many compactifications without fluxes, in particular Calabi-Yau compactifications of type II supergravity. In this paper we will lift them explicitly over a 6-torus to type IIA supergravity.

Concerning AdS compactifications, the situation is much more complicated and yet also far more interesting, since we have the AdS/CFT correspondence at our disposal. The first suggestions for AdS embeddings of Euclidean wormholes in AdS were made in~\cite{Maldacena:2004rf}, but these were not of the GS type, i.e.~sourced by axion charges. Non-axionic AdS wormholes also play a role in~\cite{Cotler:2021cqa}, where connected amplitudes are computed using constrained wormhole configurations. Despite there being no classical wormhole saddles in this context, the path integral over wormhole geometries can be done and poses problems for factorization in, for example, type IIB on $\AdS_5\times S^5$.

The nice property of GS-like wormholes is that the axions and dilatons tend to have well-understood dual CFT operators. In particular, axions and dilatons that are true moduli, in the sense that they do not appear in the potential at its minimum, correspond to exactly marginal operators. The first regular GS-like wormholes in AdS were found in~\cite{Hertog:2017owm} for the $\AdS_5\times S^5/\Z_k$ background of type IIB supergravity with $k>1$. The dual one-point functions for the operators\footnote{Here $i=1,\ldots,k$ runs over the different nodes of the dual $\N=2$ quiver CFT.} $\Tr(F_i\pm{\star F_i})^2$ was soon found to be negative~\cite{Katmadas:2018ksp}, which presents an apparent paradox. To our knowledge no other embeddings have been found. For example, the suggested proposal for an embedding inside $\AdS_3 \times S^3\times T^4 $~\cite{Arkani-Hamed:2007cpn} was shown to be singular after all~\cite{Astesiano:2022qba}, and the recent embeddings of~\cite{Marolf:2021kjc} are not of the GS type. To convincingly demonstrate that the GS wormhole embeddings of~\cite{Hertog:2017owm} constitute valid examples leading to a violation of operator positivity, we need to make sure the solutions are under control. Unfortunately, there is no full 10d description of the $\AdS_5\times S^5/\Z_k$ wormholes since the solutions were found using the corresponding 5d gauged supergravity. A proper 10d treatment would require a lift of the gauged supergravity moduli that correspond to the twisted orbifold sector, but this seems rather involved. For that reason we construct new examples in this paper that are completely explicit from a 10d point of view: we construct regular solutions for massive type IIA on $\AdS_4\times S^3\times S^3$ and for type IIB on $\AdS_5\times T^{1,1}$. Only the second has a known gauge theory dual (the Klebanov-Witten $\N=1$ quiver), and we will confirm that the constructed GS-like wormholes indeed violate operator positivity. 

Another motivation for our wormhole embeddings is the distance conjecture~\cite{Ooguri:2006in}. One can show that GS-like wormholes are only regular when the scalars traverse a large-enough distance in field space~\cite{Arkani-Hamed:2007cpn}. This distance is not parametrically super-Planckian, yet it flirts with the Planckian sizes as we will show. For that reason, one could expect that the 10d picture reveals large corrections to the lower-dimensional solutions. However, we will see that this is not the case since the solutions can be shown to be parametrically in the supergravity limit on the condition that one allows for large enough axion (dipole) charges and thus large enough wormholes. 

\smallskip

The remainder of this paper is organized as follows.
In section~\ref{sec:GSreview} we review GS wormholes in AdS and their regularity conditions.
In section~\ref{sec:flatspace_WHs} we make explicit the uplift of flat space wormholes in type IIA supergravity on a 6-torus.
In section~\ref{sec:AdS_WHs} we turn to the uplifts of AdS wormholes in type IIB supergravity where we construct solutions numerically and discuss implications for AdS/CFT.
Finally, we conclude in section~\ref{sec:conclusion}.
A more detailed discussion of the numerical methods used has been relegated to appendix~\ref{app:massiveIIA}, where we discuss the uplift of AdS wormholes in massive type IIA on $S^3\times S^3$. The accompanying source code is publicly available at \href{https://github.com/gloges/typeII-wormholes}{\texttt{github.com/gloges/typeII-wormholes}}~\cite{Loges_Type_II_AdS}.


\section{Regular Giddings-Strominger wormholes}
\label{sec:GSreview}

In this section we recall the generalised Giddings-Strominger wormholes and their associated regularity condition pioneered in~\cite{Arkani-Hamed:2007cpn}. Consider the following Euclidean action in $d\geq3$ dimensions:
\begin{equation}
    S = \frac{1}{2\kappa_d^2}\int\left({\star(\mathcal{R}-2\Lambda)} - \frac{1}{2}G_{ij}(\varphi)\d{\varphi^i}\wedge{\star\d{\varphi^j}} \right)\,.
\end{equation}
A simple set of solutions with $\O(d)$ symmetry take the form
\begin{equation}\label{eq:GSsolution}
\begin{aligned}
    \d{s^2} &= f(r)^2\,\d{r^2} + a(r)^2\,\d{\Omega_{d-1}^2} \,,\\
    \left( \frac{a'}{f} \right)^2 &= 1 + \frac{a^2}{\ell^2} + \frac{\c}{2(d-1)(d-2)a^{2d-4}} \,,\\
    \c &= G_{ij}(\varphi)\der{\varphi^i}{h}\der{\varphi^j}{h} \,,
\end{aligned}
\end{equation}
where $h(r)$ is a harmonic function, normalized to $h'=f/a^{d-1}$ so that ${\star\d{h}}=\vol_{d-1}$, and $\Lambda=-(d-1)(d-2)\ell^{-2}$. The scalars trace out a geodesic curve on the target space: the constant $\c$ is the geodesic velocity and the harmonic function $h$ plays the role of the affine parameter along the geodesic.

Wormhole geometries have $\c<0$ and correspond to time-like geodesics on the target space. In Euclidean spacetime signature the metric on the target space has indefinite signature with the time-like directions being axion directions. The above wormhole metrics are always regular, but the expressions for the geodesic curves (i.e.\ the axion and dilaton profiles) need not be. For example, a consistent truncation of Euclidean type IIB supergravity compactified over $S^5$ gives an action of the above form with the following kinetic term for the axio-dilaton:
\begin{equation}
    G_{ij}(\varphi)\partial{\varphi^i}\partial{\varphi^j} = (\partial\phi)^2 - e^{2\phi}(\partial\chi)^2 \,.
\end{equation}
The corresponding wormholes have singular axion and dilaton profiles~\cite{Bergshoeff:2005zf}.  The reason for this was understood in~\cite{Arkani-Hamed:2007cpn} in a general fashion and requires the notion of the geodesic length $D$ for the time-like geodesics, which can be computed using
\begin{equation}
\begin{aligned}
    D_d\big(\tfrac{a_0}{\ell}\big) &= \int^{\text{right}}_{\text{left}}\!\!\!\!\sqrt{-\c}\,\d{h} = 2\times\int_{a_0}^\infty\frac{\sqrt{-\c}\,\d{a}}{a^{d-1}\sqrt{1 + \frac{a^2}{\ell^2} + \frac{\c}{2(d-1)(d-2)a^{2d-4}}}}\\
    &= \sqrt{8(d-1)(d-2)}\sqrt{1+\tfrac{a_0^2}{\ell^2}}\int_1^\infty\frac{\d{\xi}}{\xi^{d-1}\sqrt{1 + \frac{a_0^2}{\ell^2}\xi^2 - \big(1+\frac{a_0^2}{\ell^2}\big)\frac{1}{\xi^{2d-4}}}} \,.
\end{aligned}
\end{equation}
The geodesic speed $\c$ has been traded in favor of the wormhole size $a_0$, which is the radius of the cross-sectional $S^{d-1}$ at the wormhole neck and is the largest real root of the equation $a'=0$ in~\eqref{eq:GSsolution}. This geodesic length is largest in the flat space limit, $\ell\to\infty$, and becomes smaller when the wormhole size is large in AdS units, $a_0\gg\ell$:
\begin{equation}
    \pi\sqrt{\frac{2(d-1)}{d-2}} = D_d(0) \geq D_d\big(\tfrac{a_0}{\ell}\big) \geq D_d(\infty) = \pi\sqrt{\frac{2(d-2)}{d-1}} \,.
\end{equation}
It is now easy to state the simple criterion which characterizes exactly when regular GS wormholes exist: there must exist a time-like geodesic for the target space $\d{s^2} = G_{ij}(\varphi)\,\d{\varphi^i}\d{\varphi^j}$ which is \emph{longer} than $D_d(\frac{a_0}{\ell})$. For an axio-dilaton pair with coupling $\beta$,
\begin{equation}
    G_{ij}(\varphi)\,\d{\varphi^i}\d{\varphi^j} = \d{\phi^2} - e^{\beta\phi}\,\d{\chi^2} \,,
\end{equation}
the longest time-like geodesic is of length $\frac{2\pi}{|\beta|}$ and the flat space condition reads
\begin{equation}
    \frac{1}{\beta^2} > \frac{d-1}{2(d-2)} \qquad (\ell\to\infty) \,.
\end{equation}
With several decoupled axio-dilaton pairs the geodesic length can be made larger by traversing a ``diagonal'' direction and the regularity condition is weakened to
\begin{equation}\label{eq:regularityflat}
    \sum_i\frac{1}{\beta_i^2} > \frac{d-1}{2(d-2)} \qquad (\ell\to\infty) \,.
\end{equation}
In section~\ref{sec:AdS_WHs} we will encounter AdS compactifications for which there is a scalar potential, in which case the solutions are no longer described by geodesics in the target space. Nevertheless, these GS solutions and their simple regularity criterion can provide some intuition for determining whether wormhole solutions exist in a more general setting.


\section{Lifting flat space wormholes}
\label{sec:flatspace_WHs}

\subsection{The reduction}
\label{sec:IIA_reduction}

Consider IIA supergravity compactified on a 6-torus. We will only require the following subsector of the 10d IIA action (in 10d Einstein frame) involving metric, dilaton and RR 4-form flux:
\be
S = \frac{1}{2\kappa_4^2}\int\d[4]{x}\,\sqrt{|g|}\left(\mathcal{R}-\frac{1}{2}(\partial\phi)^2 -\frac{1}{2\cdot4!}e^{\phi/2}F_4^2\right)\,.
\ee
This action can describe D2-branes, for instance. Indeed, the wormholes we will construct will turn out to be regular ``over-extremal" deformations of the following SUSY intersection of Euclidean D2-branes wrapping 3-cycles on the 6-torus:
\begin{equation}
\begin{aligned}
    & \times \times \times - - - \\
    & \times - - \times \times - \\
    & - \times - - \times \times \\
    & - - \times \times - \times 
\end{aligned}
\end{equation}
Below, we present an ansatz for a consistent truncation of the 6-torus reduction that is inspired by this intersection in the sense that it can at least capture the SUSY intersection. In 4d language this intersection is described by a supersymmetric instanton solution sourced by four axio-dilaton pairs. The reduction ansatz to 4d Einstein frame is
\begin{align}
    \d{s^2_{10}} &= e^{2a\varphi} \d{s^2_4} + e^{2b\varphi}\mathcal{M}_{mn}\d{\theta^m} \d{\theta^n}\,,\\
    C_3 &= \chi_1\,\d{\theta^1}\wedge\d{\theta^2}\wedge\d{\theta^3} + \chi_2\, \d{\theta^1}\wedge\d{\theta^4}\wedge\d{\theta^5} + \chi_3\,\d{\theta^2}\wedge \d{\theta^5}\wedge \d{\theta^6} + \chi_4\, \d{\theta^3}\wedge \d{\theta^4}\wedge \d{\theta^6}\,. \notag
\end{align}
The $\theta^n$ with $n=1,\ldots,6$ are the torus coordinates, $\varphi$ is the 4d radion field describing the torus volume, $\mathcal{M}_{mn}$ is a matrix of 4d fields describing the torus metric and has to have unit determinant and be both positive and symmetric. The $\chi_i$ with $i=1,\ldots,4$ are four axion fields in 4d.  For our purposes it suffices to consider a diagonal torus metric,
\be
\mathcal{M}=\operatorname{diag}\!\left(e^{\vec{\beta}_1\cdot\vec{\Phi}}, e^{\vec{\beta}_2\cdot\vec{\Phi}},\ldots, e^{\vec{\beta}_6\cdot\vec{\Phi}}\right) \,.
\ee
The $\vec{\beta}_i$ are six vectors in a 5d space and hence there are five scalars in $\vec{\Phi}$. The condition that the metric has unit determinant implies that 
$\sum_i\vec{\beta}_i=0$. 

In total we count 11 scalars in 4d in our truncation: the dilaton $\phi$, the radion $\varphi$, five torus moduli $\Phi_1,\dots,\Phi_5$ and four axions $\chi_1,\ldots,\chi_4$. This truncation of the $\mathrm{E}_{7(7)}/\SU(8)$ coset down to these 11 scalars turns out to be consistent. Once we go to Euclidean spacetime the coset signature becomes non-Euclidean meaning that the coset isotropy group is not $\SU(8)$ but some non-compact subgroup of $\mathrm{E}_{7(7)}$. Let us now fix the numbers $a,b$ and the vectors $\vec{\beta}_i$. To get the radion in canonically normalised form in 4d, we require:
\be
a^2 =\frac{3}{16}\,,\qquad a =-3b\,.
\ee 
The kinetic term for the $\mathcal{M}$-moduli becomes
\be
\frac{1}{4}\text{Tr}\!\left(\partial \mathcal{M}\partial\mathcal{M}^{-1}\right) \,,
\ee
and if the vectors $\vec{\beta}_i$ obey
\be
\vec{\beta}_i\cdot \vec{\beta}_j =2\delta_{ij}-\frac{1}{3}
\ee
then we also find canonical kinetic terms for the $\Phi$ fields; we never actually need to make an explicit choice for the vectors. With these choices the action in 4d then becomes
\begin{align}
    S = \frac{1}{2\kappa_4^2}\int \d[4]{x}\,\sqrt{|g|}\left(\mathcal{R} -\frac{1}{2}(\partial \phi)^2 -\frac{1}{2}(\partial \varphi)^2 -\frac{1}{2}\sum_{a=1}^5(\partial \Phi_a)^2+\mathcal{L}_{\text{axions}} \right) \,,
\end{align}
where 
\begin{align}
    \mathcal{L}_{\text{axions}} = -\frac{1}{2} e^{\phi/2-6b\varphi}\Big[ & e^{-(\vec{\beta}_1+\vec{\beta}_2+\vec{\beta}_3)\cdot\vec{\Phi}}(\partial \chi_1)^2 + e^{-(\vec{\beta}_1+\vec{\beta}_4+\vec{\beta}_5)\cdot\vec{\Phi}}(\partial \chi_2)^2 \nonumber \\
    & + e^{-(\vec{\beta}_2+\vec{\beta}_5+\vec{\beta}_6)\cdot\vec{\Phi}}(\partial \chi_3)^2 + e^{-(\vec{\beta}_3+\vec{\beta}_4+\vec{\beta}_6)\cdot\vec{\Phi}}(\partial \chi_4)^2 \Big]\,.
\end{align}
If one groups the seven scalars $\phi,\varphi,\Phi_1,\ldots,\Phi_5$ into a vector $\vec{t}$ then the axion kinetic terms can be rewritten as
\be
\L_\text{axions} = -\frac{1}{2}\sum_{i=1}^4 e^{\vec{\alpha}_i \cdot\vec{t}} (\partial \chi_i)^2\,, 
\ee
where the $\vec{\alpha}_i$ are four orthogonal vectors of length two: $\vec{\alpha}_i\cdot\vec{\alpha}_j = 4\delta_{ij}$.
Hence we can construct a new base of canonically normalised scalars as
\be
s_i = \frac{1}{2}\vec{\alpha}_i\cdot \vec{t}
\ee
for $i=1,\ldots,4$ and with $s_5, s_6, s_7$ orthogonal to them. Then the action becomes:
\begin{align}
    S = \frac{1}{2\kappa_4^2}\int \d[4]{x}\,\sqrt{|g|}\left(\mathcal{R} -\frac{1}{2}\sum_{i=1}^4\left[ (\partial s_i)^2 +e^{2s_i}(\partial \chi_i)^2\right] -\frac{1}{2}\sum_{k=5}^7(\partial s_k)^2 \right) \,.
\end{align}
Upon going to Euclidean signature the axion kinetic terms effectively flip sign, or alternatively, we can describe them by tensors obtained from $F_6$ reduction.
We will further truncate $s_5, s_6$ and $s_7$ such that our final truncation comprises eight fields that combine into four decoupled axio-dilaton pairs, each with coupling $\beta_i=2$. The condition for regularity in 4d in~\eqref{eq:regularityflat} is satisfied, since we have
\be
\sum_{i=1}^4\frac{1}{\beta_i^2} = 1 > \frac{3}{4} = \left(\frac{D_4(0)}{2\pi}\right)^2 \,.
\ee
Therefore we conclude that there exist regular GS wormholes. It is critical that there were (at least) four decoupled axio-dilaton pairs so as to exceed the $3/4$ lower bound.

We can make a further truncation and work with the single effective axio-dilaton pair that has $\beta_{\text{eff}}=1$. Clearly this requires some isotropy and corresponds to giving all of the axions the same charge. Once one does this, one needs to set $\Phi_a=0$ in order for this to be consistent, which is possible exactly for isotropic axion charges. Then it is easy to see why the effective $\beta$ is one, since the axion kinetic terms are
\be
\mathcal{L}_{\text{axions}} = -\frac{1}{2} e^{\phi/2-6b\varphi} (\partial\chi)^2\equiv -\frac{1}{2} e^s(\partial\chi)^2\,,
\ee
where now $\chi \equiv 2\chi_1 =2\chi_2=2\chi_3=2\chi_4$. The effective dilaton coupling is one since the vector $(1/2, {-6}b)$ has length one. The orthogonal scalar combination,
\be\label{eq:orthog}
6b\phi+\frac{\varphi}{2} \,,
\ee
can then be put to zero since it decouples. This means that everywhere $e^{6b\phi}=e^{-\varphi/2}$ or equivalently $e^{\phi/2}=e^{-2b\varphi}$. 
This is the $\beta=1$ dilaton coupling that~\cite{Bergshoeff:2004pg, Davidse:2003ww} claimed to find in general CY compactifications in the type IIA RR sector.

\subsection{Lifting the extremal solutions}

The 4d extremal solutions read
\be
\d{s_4^2} = \d{r^2} + r^2\d{\Omega_3^2} \,, \qquad e^{s/2} = H(r)\,,\qquad \chi = \frac{2}{H(r)}\,,
\ee
where $H(r)$ is a harmonic function of $r$. A 10d BPS intersection of Euclidean D2s with four charges for the four different axions would give the following string frame metric and string dilaton profile:
\begin{equation}
\begin{aligned}
    \d{s^2} &= \sqrt{H_1H_2H_3H_4}\left(\d{r^2} + r^2\,\d{\Omega_3^2}\right) + \sqrt{\frac{H_3H_4}{H_1H_2}}\,\d{\theta_1^2} + \sqrt{\frac{H_2H_4}{H_1H_3}}\,\d{\theta_2^2} \\
    &\qquad + \sqrt{\frac{H_2H_3}{H_1H_4}}\,\d{\theta_3^2}+\sqrt{\frac{H_1H_3}{H_2H_4}}\,\d{\theta_4^2}+ \sqrt{\frac{H_1H_4}{H_2H_3}}\,\d{\theta_5^2}+\sqrt{\frac{H_1H_2}{H_3H_4}}\,\d{\theta_6^2} \,,\\
    e^\phi &= (H_1H_2H_3H_4)^{1/4} \,.
\end{aligned}
\end{equation}
Now let us look at the ``isotropic'' solution for which all four charges are equal and thus all four harmonics are equal,
\be
H\equiv H_1=H_2=H_3=H_4\,,
\ee
in which case the 10d solution simplifies to
\begin{equation}\label{eq:intersection}
\begin{aligned}
    \d{s^2} &= H^2\left(\d{r^2} + r^2\,\d{\Omega_3^2}\right) + \d{\theta_1^2} + \d{\theta_2^2}+\d{\theta_3^2}+\d{\theta_4^2}+ \d{\theta_5^2}+\d{\theta_6^2} \,,\\
    e^\phi &= H \,.
\end{aligned}
\end{equation}
Let us see whether our lift indeed reproduces this. To respect isotropy we put all $\Phi_a=0$ and then $s$ with $\beta$-value equal to $1$ is given by $e^s = e^{\phi/2-6b\varphi}$ and the orthogonal combination vanishes,
$6b\phi + \varphi/2=0$. This means that
\be
e^{\phi/2}=e^{-2b\varphi}\qquad \text{and}\qquad e^s=e^{2\phi}\,,
\ee
and we get the result~\eqref{eq:intersection} on the nose from uplifting.

\subsection{Lifting the wormhole}
For the wormhole we have
\be
\d{s^2_4} = f(r)^2\,\d{r^2} + a(r)^2\,\d{\Omega_3^2} \,.
\ee
with $f(r)$ a gauge choice and $a(r)$ the scale factor. We will use the following form of the wormhole solution,
\begin{equation}
\begin{aligned}
    \d{s^2} &= \left(1 + \frac{a_0^4}{4r^4}\right)\left(\d{r^2} + r^2\,\d{\Omega_3^2}\right)\,,\\
    e^{s/2} &= A\sin\!\left[\sqrt{3}\arctan\left(\frac{a_0^2}{2r^2}\right)+B\right]\,,\\
\end{aligned}
\end{equation}
with $A, B$ fixed by boundary conditions and $a_0$ the radius of the wormhole neck. The wormhole's left-hand side is at $r=0$ while the right-hand side is at $r=\infty$; one can check that in both of these asymptotic regions the metric is flat. Since arctan interpolates between $\pi/2$ and $0$ from the left side to the right, for values of $B$ that are in the range
\be
0<B<\left(1 - \tfrac{\sqrt{3}}{2}\right)\pi \approx 0.42 \,,
\ee
the functional form of $e^{s/2}$ remains positive everywhere. So with a value of $B$ somewhere in the middle of the interval, a value of $A$ that is small enough and a value of $a_0$ that is large enough, the 4d metric is classical and the ``coupling'' $e^s$ is small everywhere.  

To uplift we will be maximally safe and not put the free combination $6b\phi + \varphi/2$ to zero, but to a constant. This means that
\be
e^{\phi/2} = C^2e^{-2b\varphi}\qquad \text{and}\qquad e^\phi = C^3e^{s/2} \,,
\ee
with $C$ some arbitrary positive constant.
The uplift formula is simply\footnote{To lift the axions to the gauge field in 10d, we used that
$\chi_1=\chi_2=\chi_3=\chi_4=\chi/2$ with $$\chi=\frac{2}{A}\cot\!\left[\sqrt{3}\arctan\left(\frac{a_0^2}{2r^2}\right)+ B\right]$$.}:
\begin{equation}
\begin{aligned}
    \d{s^2} &= A^2\sin^2\!\left[\sqrt3 \arctan\left(\frac{a_0^2}{2r^2}\right)+B\right]\left(1 + \frac{a_0^4}{4r^4}\right)\left(\d{r^2} + r^2\,\d{\Omega_3^2}\right) + C^2\sum_{i=1}^6\d{\theta_i^2}\,,\\
    e^\phi &= C^3 A\sin\!\left[\sqrt3 \arctan\left(\frac{a_0^2}{2r^2}\right)+B\right]\,,\\
    C_3 &= \frac{i}{A}\cot\!\left[\sqrt3 \arctan\left(\frac{a_0^2}{2r^2}\right)+B\right]\left(\d{\theta^{123}} + \d{\theta^{145}} + \d{\theta^{256}} + \d{\theta^{346}}\right)\,.
\end{aligned}
\end{equation}
We can always take $C$ large to get a large-enough internal torus in string frame. Since there is always a value of $r$ at which $e^\phi=C^3A$, we must also take $A$ small to maintain weak coupling throughout, but then the warp factor in front of the 4d metric na\"ively becomes small as well. However, we also have the parameter $a_0$ available to adjust, and its effect is most clear after rescaling the radial coordinate via $r=a_0\rho/\sqrt{2}$:
\begin{equation}
\begin{aligned}
    \d{s^2} &= \frac{A^2a_0^2}{2}\sin^2\!\left[\sqrt3 \arctan\left(\frac{1}{\rho^2}\right)+B\right]\left(1 + \frac{1}{\rho^4}\right)\left(\d{\rho^2} + \rho^2\,\d{\Omega_3^2}\right) + C^2\sum_{i=1}^6\d{\theta_i^2}\,,\\
    e^\phi &= C^3 A\sin\!\left[\sqrt3 \arctan\left(\frac{1}{\rho^2}\right)+B\right]\,,\\
    C_3 &= \frac{i}{A}\cot\!\left[\sqrt3 \arctan\left(\frac{1}{\rho^2}\right)+B\right]\left(\d{\theta^{123}} + \d{\theta^{145}} + \d{\theta^{256}} + \d{\theta^{346}}\right)\,.
\end{aligned}
\end{equation}
This makes it explicit that the wormhole geometry in 10d geometry can be weakly coupled and weakly curved, so it appears to be under full control.

Another possibility is to choose integration constants such that the string coupling is large throughout the flow. Then one has to make sure that the 11d curvature of the uplifted solution is small.
Again, with the proper tuning of $A, B, C$ and $a_0$ this can easily be achieved. Note also that we can T-dualise these solutions to Euclidean type IIB with intersecting D3/D1-branes on a six-torus in the following manner:
\begin{equation}
\begin{aligned}
    \times \times \times - - \times\\
    \times - - \times \times \times \\
    - \times - - \times - \\
    - - \times \times - - 
\end{aligned}
\end{equation}
Finally, note that if we were to write the solution using $C_5$ instead of $C_3$ then all fields would be real.

\subsection{Wick rotation}
Since a stack of parallel Euclidean D2-branes can be Wick-rotated to a stack of Lorentzian D2-branes, one can wonder whether we have found a contradiction with the non-existence of over-extremal D-branes, since the Wick rotation of our wormhole would be an intersection of Lorentzian D2-branes with total tension smaller than charge, which cannot exist (i.e.\ be without naked singularities, let alone regular). This is, however, not the case, and the reason is that our solutions involve \emph{intersecting} branes. For example, consider Wick rotating the $\theta_1$ coordinate. This would make the D2-branes, wrapping directions 123 and 145 Lorentzian, but the others remain Euclidean, with ``wrong sign" $F_4$ energies. Hence we find no contradiction; this heavily relies upon the separate axio-dilaton pairs having too-large couplings, such that we were forced to invoke multiple axions and hence multiple intersecting Euclidean branes.


\section{Lifting AdS wormholes}
\label{sec:AdS_WHs}

In this section we construct regular AdS wormholes in type IIB. The new solutions necessarily involve models in which the dilaton partners of the axions are not massless so that the geodesic motion is distorted. In such a case there is no easy regularity criterion as for geodesics. A recent simple bottom-up analysis~\cite{Andriolo:2022rxc}, related to older work~\cite{Kallosh:1995hi}, studies the possible wormhole solutions for an axion with a massive dilaton of the following kind
\be
e^{-1}\mathcal{L} =\mathcal{R}-\tfrac{1}{2}(\partial\phi)^2 +\tfrac{1}{2}e^{\beta\phi}(\partial\chi)^2 - V(\phi)\,,
\ee
where
\be
V = \Lambda +m^2\phi^2 + \mathcal{O}(\phi^3)\,.
\ee
In~\cite{Andriolo:2022rxc} it was found (for $\Lambda=0$) that if $\beta$ is too large to allow for a smooth wormhole in the absence of $V(\phi)$, then there is at least a critical mass above which the wormhole exists. After all, in the large-$m$ limit, the dilaton can be integrated out to obtain the pure axion wormhole, which is always regular. It is likely that this critical mass depends on the size of the wormhole neck and hence the axion charge. When $m^2$ is of the order of $\Lambda$ it is unclear what happens, especially when there is more than a single axio-dilaton pair present.

We start first by recalling the AdS embeddings of Euclidean wormholes pioneered in~\cite{Hertog:2017owm} and comment on their relation with the distance conjecture.


\subsection{Type IIB on \texorpdfstring{$S^5/\Z_k$}{S5/Zk}}
It was noticed in~\cite{Hertog:2017owm} that the problem of fitting regular wormholes into the well-known $\AdS_5\times S^5$ background could be alleviated by orbifolding to $\AdS_5\times S^5/\Z_k$. The orbifolding generates extra moduli and the effective geodesic distance can be enlarged in a similar way to what was envisaged in~\cite{Arkani-Hamed:2007cpn}. The dual description is the so-called $\N=2$ necklace quiver with $k$ nodes. The classification of supersymmetric and non-supersymmetric geodesics was then performed in~\cite{Ruggeri:2017grz} and an interpretation in terms of the dual CFT was carried out in~\cite{Katmadas:2018ksp}. This analysis revealed the problematic nature of the regular wormholes, already alluded to in~\cite{Bergshoeff:2005zf}: the wormholes violate the dual BPS bound since
\be
\langle \Tr(F_i-\star F_i)^2\rangle < 0\,.
\ee
Here $i$ runs over the gauge nodes from $1,\ldots,k$ and $F_i$ is the Yang-Mills field strength tensor of the $i^\text{th}$ node. Although the operator is manifestly positive, the holographic computation of the dual one-point function (vev) fails to respect it. This is simply a manifestation that in gravity we can violate BPS bounds, but not in field theories and wormholes are therefore problematic since they correspond to over-extremal instantons. One would hope that the gravity computation somehow signals the sickness, yet the solution is completely smooth inside the $\AdS_5$ factor. What happens inside the orbifolded 5-sphere is less clear since there are already the orbifold singularities and the flow of the scalars happens in the twisted sector. Hence the problem of lifting the wormholes is not so easy and we refrain from attempting this. Instead in the next subsection we discuss the closely related solution of type IIB on $T^{1,1}$ which is related by an RG flow to the $S^5/\Z_2$ theory and has the benefit of being completely explicit as there is no orbifolding. But before we do so we want to comment on the distance conjecture. 

For this purpose we need some details about the wormhole solutions. The sigma model of the orbifolded setup is given by $\SL(k+1,\R)/\GL(k,\R)$ in the Euclidean version of the supergravity theory. The corresponding metric on this coset space (i.e.\ kinetic term) reads
\be
\d{s^2} = 4\,\d{U^2} - e^{4U}\N^2 + e^{-2U}\sum_{i=1}^{k-1}\big(\d{\zeta_i^2} - \d{\tilde{\zeta}_i^2}\big)\ee
with
\be
\N=\d{a} + \frac{1}{2}\sum_{i=1}^{k-1} \big(\zeta_i\,\d{\tilde{\zeta}_i} - \tilde{\zeta}_i\, \d{\zeta_i}\big) \,.
\ee
So we have $2k$ real variables, $U, a, \zeta_i, \tilde{\zeta}_i$, of which $a$ and $\tilde{\zeta}_i$ are the ``Wick-rotated axions.'' To cut a long story short~\cite{Hertog:2017owm, Ruggeri:2017grz}, one can show that the regular geodesics can be brought, by means of an isometry transformation, to a simple geodesic of an $\SL(2, \R)/\SO(1,1)$ subspace formed by a single axio-dilaton pair with coupling $\beta=1$. In this field basis all the extra $2(k-1)$ fields are constant. The surprising fact is that the geodesic length, in string units, remains fixed with respect to $k$, but in Planckian units it is not. In 5d we have the relation
\be
M_\text{Pl}^3 = M_\text{s}^3g_\text{s}^{-2}\, \text{Vol}_5\,,
\ee
where $M_\text{s}$ is the string scale and Vol$_5$ is the volume of the internal space in string units. Note that this volume goes down with a factor $k$. So the field distance along the geodesic in moduli space grows with a factor $k^{1/3}$ in Planck units and can therefore be made arbitrary large, indicating an apparent violation of the distance conjecture. However, note that our system is such that the number of massless degrees of freedom scales with $k$ so that in terms of the species scale the distance remains constant. This is therefore consistent with the most na\"ive formulation of the distance conjecture for Euclidean theories. 

We want to emphasize there is yet no clear definition of the distance conjecture in Euclidean theories because the target space metric is not positive definite.\footnote{A similar subtlety arises in formulating the AdS distance conjecture~\cite{Lust:2019zwm} with metric spaces whose metrics are not positive definite. This subtlety motivated a recent study~\cite{Shiu:2022oti} of distances traversed by scalar fields interpolating between flux vacua.} We will be agnostic about this and have a pragmatic view in which we simply compute the absolute value of the geodesic distance, but it could very well be that only the distance in the dilaton space is to be considered for the distance conjecture. The reason is that axion momenta are really more like quantised fluxes and, when dualised to higher forms, would not be considered in the standard definition of the target space distance.

\subsection{Type IIB on \texorpdfstring{$T^{1,1}$}{T11}}
\label{sec:wh_typeIIB}

In what follows we consider a consistent truncation of type IIB supergravity on $T^{1,1}$. This model is closely connected to the one in the previous section but has the benefit of being fully explicit in 10d since there is no orbifolding. The connection with the dual CFT was studied first in~\cite{Klebanov:1998hh}.
As for instance explained in~\cite{Louis:2016msm}, the moduli space of $\AdS_5\times T^{1,1}$ has real dimension equal to ten; what we will consider below is a consistent truncation consisting of four moduli and two massive dilatons. Hence the wormhole solutions are possible because of the negative axion energy-momentum, but are not exactly of GS type due to the dilaton masses.

\subsubsection{The reduction}
The 10d Einstein frame equations of motion of the bosons in type IIB supergravity are recalled in appendix~\ref{app:typeII_eoms}. 
Type IIB supergravity has an $\AdS_5\times T^{1,1}$ solution with $8$ supercharges supported only by $F_5$-flux~\cite{Klebanov:1998hh}:
\begin{equation}
\begin{aligned}
    \d{s_{10}^2} &= \ell^2\left(\frac{\d{r^2}}{1+r^2} + r^2\,\d{\Omega_4^2}\right) + \ell^2\left(\d{s_\text{KE}^2} + \eta^2\right) \,,\\
    e^\Phi &= g_\text{s} \,,\\
    H_3 &= F_1 = F_3 = 0 \,,\\
    F_5 &= 4\ell^4(1 - i\star)\vol_{T^{1,1}} \,.
\end{aligned}
\end{equation}
The factor of $i$ appears because of the Wick rotation to Euclidean signature.
The metric on $T^{1,1}$ has been written emphasizing its structure as a $\U(1)$ fibration over an $S^2\times S^2$ K\"ahler-Einstein base.\footnote{Explicitly,
\begin{equation*}
    \d{s_\text{KE}^2} = \frac{1}{6}\sum_{i=1,2}\big(\d{\vartheta_i^2} + \sin^2{\vartheta_i}\,\d{\varphi_i^2}\big) \,, \qquad \eta = \frac{1}{3}\big(\d{\psi} + \cos{\vartheta_1}\,\d{\varphi_1} + \cos{\vartheta_2}\,\d{\varphi_2} \big) \,,
\end{equation*}
with $\vartheta_i\in[0,\pi]$, $\varphi_i\in[0,2\pi)$ and $\psi\in[0,4\pi)$. The second cohomology is spanned by
\begin{equation*}
    \Phi_2 = \frac{1}{\sqrt{72}}\big(\sin{\vartheta_1}\,\d{\vartheta_1}\wedge\d{\varphi_1} - \sin{\vartheta_2}\,\d{\vartheta_2}\wedge\d{\varphi_2}\big)
\end{equation*}
and the third cohomology is spanned by $\Phi_2\wedge\eta$.
}
For the sake of finding wormhole geometries in this background we use the following ansatz,
\begin{equation}\label{eq:typeIIB_T11_ansatz}
\begin{aligned}
    \d{s_{10}^2} &= \ell^2e^{-\frac{2}{3}(4u+v)}\left[f(r)^2\,\d{r^2} + q(r)^2\,\d{\Omega_4^2}\right] + \ell^2\big(e^{2u}\,\d{s_\text{KE}^2} + e^{2v}\eta^2\big) \,,\\
    e^\Phi &= g_\text{s}e^\phi \,,\\
    B_2 &= \ell^2g_\text{s}^{1/2}b\,\Phi_2 \,,\\
    C_0 &= ig_\text{s}^{-1}\chi \,,\\
    C_2 &= i\ell^2g_\text{s}^{-1/2}c\,\Phi_2 \,,\\
    F_5 &= 4\ell^4(1-i\star)\vol_{T^{1,1}} \,,
\end{aligned}
\end{equation}
where $\Phi_2$ spans the second cohomology of $T^{1,1}$ (normalized to $|\Phi_2|^2=1$) and $\phi\to0$ at the boundary. The non-trivial 5d equations of motion are independent of $\ell$ and $g_\text{s}$ by design, and can be shown to be derived from the following 5d action~\cite{Cassani:2010na}:
\begin{equation}\label{eq:5Daction}
\begin{aligned}
    S_5 &= \frac{1}{2\kappa_5^2}\int\Big({\star \mathcal{R}} - \frac{1}{2}\d{\phi}\wedge{\star\d{\phi}} - \frac{1}{2}e^{-4u-\phi}\d{b}\wedge{\star\d{b}} \\
    &\qquad\qquad\qquad + \frac{1}{2}e^{2\phi}\d{\chi}\wedge{\star\d{\chi}} + \frac{1}{2}e^{-4u+\phi}(\d{c}-\chi\,\d{b})\wedge{\star(\d{c}-\chi\,\d{b})}\\
    &\qquad\qquad\qquad - \frac{28}{3}\d{u}\wedge{\star\d{u}} - \frac{8}{3}\d{u}\wedge{\star\d{v}} - \frac{4}{3}\d{v}\wedge{\star\d{v}} - {\star\mathcal{V}}\Big) \,,\\
    \mathcal{V} &= 2e^{-\frac{8}{3}(4u+v)}\left(2e^{4u+4v} - 12e^{6u+2v} + 4\right) \,.
\end{aligned}
\end{equation}
The minimum of the potential, $\mathcal{V}=-12$, occurs at $u=v=0$. The associated equations of motion have solutions in 5d AdS units with $g_s$ scaled out and so one should appeal to the 10d uplift formula for a full understanding.

\subsubsection{Symmetries and axion charges}

There are three shift symmetries,
\begin{equation}
\begin{aligned}
    \delta_1\chi &= \lambda_1 \,, & \qquad \delta_2\chi &= 0 \,, & \qquad \delta_3\chi &= 0 \,,\\
    \delta_1b &= 0 \,, & \delta_2b &= \lambda_2 \,, & \delta_3b &= 0 \,,\\
    \delta_1c &= \lambda_1b \,, & \delta_2c &= 0 \,, & \delta_3c &= \lambda_3 \,,
\end{aligned}
\end{equation}
(with $\lambda_i$ constant) which descend from the gauge symmetries corresponding to $C_0$, $C_2$ and $B_2$, respectively. Note that $\d{c}$ ($\d{C_2}$) transforms under the $\chi$ ($C_0$) shift, but $\d{c}-\chi\,\d{b}$ (${F_3=\d{C_2}-C_0\wedge H_3}$) is gauge-invariant. The $\delta_1$ transformation is an element of $\SL(2,\R)$, under which the fields transform as
\begin{equation}
    e^{-\phi} \mapsto e^{-\phi} \,, \qquad \chi\mapsto \chi + \lambda_1 \,, \qquad    
    \begin{bmatrix}
        b\\ c
    \end{bmatrix} \mapsto \begin{bmatrix}
        1 & 0\\ \lambda_1 & 1
    \end{bmatrix}\begin{bmatrix}
        b\\ c
    \end{bmatrix}\,.
\end{equation}
There is also the rescaling symmetry of $\SL(2,\R)$, for which
\begin{equation}\label{eq:rescale}
    e^{-\phi} \mapsto a^2e^{-\phi} \,, \qquad \chi \mapsto a^2\chi \,, \qquad
    \begin{bmatrix}
        b\\ c
    \end{bmatrix} \mapsto \begin{bmatrix}
        a^{-1} & 0\\ 0 & a
    \end{bmatrix}\begin{bmatrix}
        b\\ c
    \end{bmatrix}\,.
\end{equation}
Since we work in the supergravity approximation where these continuous symmetries are not broken to discrete subgroups, we can use the above perturbative symmetries to simplify the numerical analysis.

The axion equations of motion can be solved in terms of constant axion charges as
\begin{equation}\label{eq:axion_charges}
\begin{aligned}
    \star\big[e^{-4u-\phi}\,\d{b} + e^{-4u+\phi}\chi(\d{c}-\chi\,\d{b})\big] &\equiv \mathfrak{q}_1\,\vol_4\,,\\
    {\star\big[e^{-4u+\phi}(\d{c}-\chi\,\d{b})\big]} &\equiv \mathfrak{q}_2\,\vol_4\,.
\end{aligned}
\end{equation}
These charges correspond to the Noether charges under the two commuting shift symmetries $\delta_2$ and $\delta_3$. From a 10d viewpoint we have
\begin{equation}
\begin{aligned}
    {\star\big(e^{-\Phi}H_3 - e^\Phi C_0\wedge F_3\big)} &\quad\sim\quad \mathfrak{q}_1\,\vol_4\wedge\eta\wedge\Phi_2 \,,\\
    {\star\big(e^\Phi F_3\big)} &\quad\sim\quad \mathfrak{q}_2\,\vol_4\wedge\eta\wedge\Phi_2 \,.
\end{aligned}
\end{equation}
These charges are conserved by the equations of motion, but $\mathfrak{q}_1$ is clearly gauge-dependent since $C_0$ appears explicitly: $\mathfrak{q}_1$ is a ``Page charge'' in the language of~\cite{Marolf:2000cb}. There is no such ambiguity for $\mathfrak{q}_2$, which is both gauge-independent and quantized. However, since $\mathfrak{q}_2$ is integer-quantized in units of $(\ell_\text{s}/\ell)^6$ and $\ell\gg\ell_\text{s}$, we are justified in taking a continuous-charge approximation and will not be bothered with quantisation conditions.

For completeness, under the shift symmetries,
\begin{equation}
\begin{aligned}
    \delta_1\mathfrak{q}_1 &= \lambda_1\mathfrak{q}_2 \,, & \qquad \delta_2\mathfrak{q}_1 &= 0 \,, & \qquad \delta_3\mathfrak{q}_1 &= 0 \,,\\
    \delta_1\mathfrak{q}_2 &= 0 \,, & \qquad \delta_2\mathfrak{q}_2 &= 0 \,, & \qquad \delta_3\mathfrak{q}_2 &= 0 \,,
\end{aligned}
\end{equation}
and under the ``rescaling symmetry,''
\begin{equation}
    \mathfrak{q}_1 \mapsto a\,\mathfrak{q}_1 \,, \qquad \mathfrak{q}_2 \mapsto a^{-1}\mathfrak{q}_2 \,.
\end{equation}
In terms of the $\mathfrak{q}_1$ and $\mathfrak{q}_2$ charges, the (remaining) 5d equations of motion take the form
\begin{equation}\label{eq:5d_equations}
\begin{aligned}
    \square\phi &= -e^{2\phi}(\partial\chi)^2 - \frac{1}{2} e^{4u}\big[ e^\phi(\mathfrak{q}_1-\mathfrak{q}_2\chi)^2 + e^{-\phi}\mathfrak{q}_2^2 \big](\partial h)^2 \,,\\
    \d{\star(e^{2\phi}\d{\chi})} &= -e^{4u+\phi}\mathfrak{q}_2(\mathfrak{q}_1-\mathfrak{q}_2\chi)(\partial h)^2 \,,\\
    \square(7u+v) &= \frac{3}{8}\partial_u\mathcal{V} - \frac{3}{4}e^{4u}\big[ e^\phi(\mathfrak{q}_1-\mathfrak{q}_2\chi)^2 - e^{-\phi}\mathfrak{q}_2^2 \big](\partial h)^2 \,,\\
    \square(u+v) &= \frac{3}{8}\partial_v\mathcal{V} \,,\\
    2R_{\mu\nu} &= \partial_\mu\phi\partial_\nu\phi - e^{2\phi}\partial_\mu\chi\partial_\nu\chi + e^{4u}\big[ e^\phi(\mathfrak{q}_1-\mathfrak{q}_2\chi)^2 - e^{-\phi}\mathfrak{q}_2^2 \big]\partial_\mu h\partial_\nu h\\
    &\qquad\qquad + \frac{56}{3}\partial_\mu u\partial_\nu u + \frac{8}{3}(\partial_\mu u\partial_\nu v + \partial_\mu v\partial_\nu u) + \frac{8}{3}\partial_\mu v\partial_\nu v - \frac{2}{3}g_{\mu\nu}\mathcal{V} \,.
\end{aligned}
\end{equation}
Here $h(r)$ is harmonic and normalized to $h'=f/q^4$ so that ${\star\d{h}}=\vol_4$. Solutions have definite parity when $\mathfrak{q}_1=0$, which we take for simplicity when constructing numerical solutions. This choice can be undone using the $\delta_1$ transformation above as long as $\mathfrak{q}_2$ is non-zero (which will be the case for the non-singular solutions of interest).

\subsubsection{Review of supersymmetric solutions}
Here we review the supersymmetric instanton solutions which appear in the Klebanov-Witten model~\cite{Imaanpur:2016beu}. These solutions serve as an anchor point to test our understanding of the truncation, axion charges and especially whether or not our wormholes violate positivity bounds, since these bounds are saturated for SUSY instantons. 

We can see from the 5d equations of motion~\eqref{eq:5d_equations} that the matter stress tensor vanishes exactly when
\begin{equation}
    \mathfrak{q}_1 - \mathfrak{q}_2\chi = \pm\mathfrak{q}_2\,e^{-\phi}
\end{equation}
with $u=v=0$ sitting at the minimum of $\mathcal{V}$. This statement is invariant under all of the symmetries (all three shift symmetries and the rescaling). In particular, recall that $\mathfrak{q}_1$ transforms under constant shifts of $\chi$. Picking the bottom sign (c.f.~\cite{Imaanpur:2016beu}),
\begin{equation}
    \chi - e^{-\phi} = \frac{\mathfrak{q}_1}{\mathfrak{q}_2} = \text{const.} \qquad\implies\qquad F_3 = -ie^{-\Phi}H_3 \,.
\end{equation}
The metric is empty AdS$_5$, and the remaining equation of motion for $\phi$ can be reorganised into the form
\begin{equation}
    \d{\star\d{\left(e^\phi + \frac{1}{2}\mathfrak{q}_2^2h^2\right)}} = 0\,.
\end{equation}
Thus we have
\begin{equation}
\begin{aligned}
    e^\phi &= A_1 + A_2h - \frac{1}{2}\mathfrak{q}_2^2h^2 \,,\\
    \chi &= \frac{\mathfrak{q}_1}{\mathfrak{q}_2} + e^{-\phi} \,,\\
    \d{b} &= -\mathfrak{q}_2\,\d{h} \,,\\
    \d{c} &= -\mathfrak{q}_1\,\d{h} \,,
\end{aligned}
\end{equation}
for $A_1,A_2$ two integration constants (more generally, the harmonic functions accompanying $\mathfrak{q}_1$, $\mathfrak{q}_2$ and $A_2$ need not be the same).

\subsubsection{Further truncations and Giddings-Strominger ``approximations''}
\label{sec:GSapprox}

There are several simplifications that can be made, some wholly unjustified, to gain some insight into whether regular wormholes exist for the system under consideration. Recall that we set $\mathfrak{q}_1=0$.

\begin{itemize}
    \item $\mathbf{\mathfrak{q}_2=0}$: If one sets $b=c=0$ by setting both of the charges to zero, then $u$ and $v$ are not sourced and $\phi,\chi$ comprise an axio-dilaton pair with $\beta=2$. The corresponding GS wormholes are always singular:
    \begin{equation}
        \left(\frac{D_5(q_0)}{2\pi}\right)^2 \geq \left(\frac{D_5(\infty)}{2\pi}\right)^2 = \frac{3}{8} > \frac{1}{\beta^2} \,.
    \end{equation}
    \item \textbf{$\mathbf{\chi=0}$ and $\mathbf{u,v}$ frozen}: If one chooses the solution $\chi=0$ and makes an \emph{ad hoc} change to the potential so that $u,v$ are frozen at the minimum of $\mathcal{V}$, then $\phi,c$ comprise an axio-dilaton pair with $\beta=1$. The corresponding GS wormholes are always smooth:
    \begin{equation}
        \frac{1}{\beta^2} > \frac{2}{3} = \left(\frac{D_5(0)}{2\pi}\right)^2 \geq \left(\frac{D_5(q_0)}{2\pi}\right)^2 \,.
    \end{equation}
    \item \textbf{$\mathbf{\chi=0}$ and $\mathbf{u,v}$ massless}: If one chooses the solution $\chi=0$ and makes an \emph{ad hoc} change to the potential so that $\mathcal{V}=-12$ and $u,v$ are massless, then the 5d action may be rewritten as
    \begin{equation}
        \frac{1}{2\kappa_5^2}\int\Big({\star(\mathcal{R}+12)} - \frac{1}{2}\delta_{ij}\d{\varphi^i}\wedge{\star\d{\varphi^j}} + \frac{1}{2}e^{\sqrt{2}\varphi^1}\d{c}\wedge{\star\d{c}}\Big) \,,
    \end{equation}
    where
    \begin{equation}
        \varphi^1 = \frac{-4u+\phi}{\sqrt{2}} \,, \qquad \varphi^2 = \frac{4u+\phi}{\sqrt{2}} \,, \qquad \varphi^3 = \frac{4(u+v)}{\sqrt{6}} \,.
    \end{equation}
    The scalars $\varphi^2$ and $\varphi^3$ decouple, so we can set $\phi=-4u=4v$, and $\varphi^1,c$ comprise an axio-dilaton pair with $\beta=\sqrt{2}$, which satisfies
    \begin{equation}
        \left(\frac{D_5(0)}{2\pi}\right)^2 = \frac{2}{3} \;>\; \frac{1}{\beta^2} \;>\; \frac{3}{8} = \left(\frac{D_5(\infty)}{2\pi}\right)^2 \,.
    \end{equation}
    This means that in this ``approximation'' the GS solution is regular only if the wormhole is large enough (${q_0>0.7727}$) and for small wormholes the scalars diverge as $(-u),v,\phi\to+\infty$ as one moves into the bulk: the effective string coupling $g_\text{s}e^\phi$ goes to $+\infty$ while the volumes of the $S^2$ and $S^3$ of $T^{1,1}$ both go to zero.
\end{itemize}
The above three (over)simplifications lead to the following picture. There is a balance between $\chi$ and $b,c$ (via $\mathfrak{q}_2$) that is needed to get smooth wormhole solutions of all sizes. If $b,c$ are set to zero then all potential wormholes are singular. If $\chi$ is set to zero then it is clear that large wormholes exist in this model; even with dramatic changes to the potential we saw that regular GS wormholes can be found: presumably the true profiles will lie somewhere between those of the $\beta=1$ and $\beta=\sqrt{2}$ GS wormholes. The existence of small wormholes, however, is less clear and must be addressed numerically. We will see that regular wormholes of any size \emph{do} exist, but the scalars inevitably grow very large in the neck region as one takes $q_0\ll 0.7727$.

\subsubsection{Numerical solutions}

\begin{figure}[p]
    \centering
    \includegraphics[width=\textwidth]{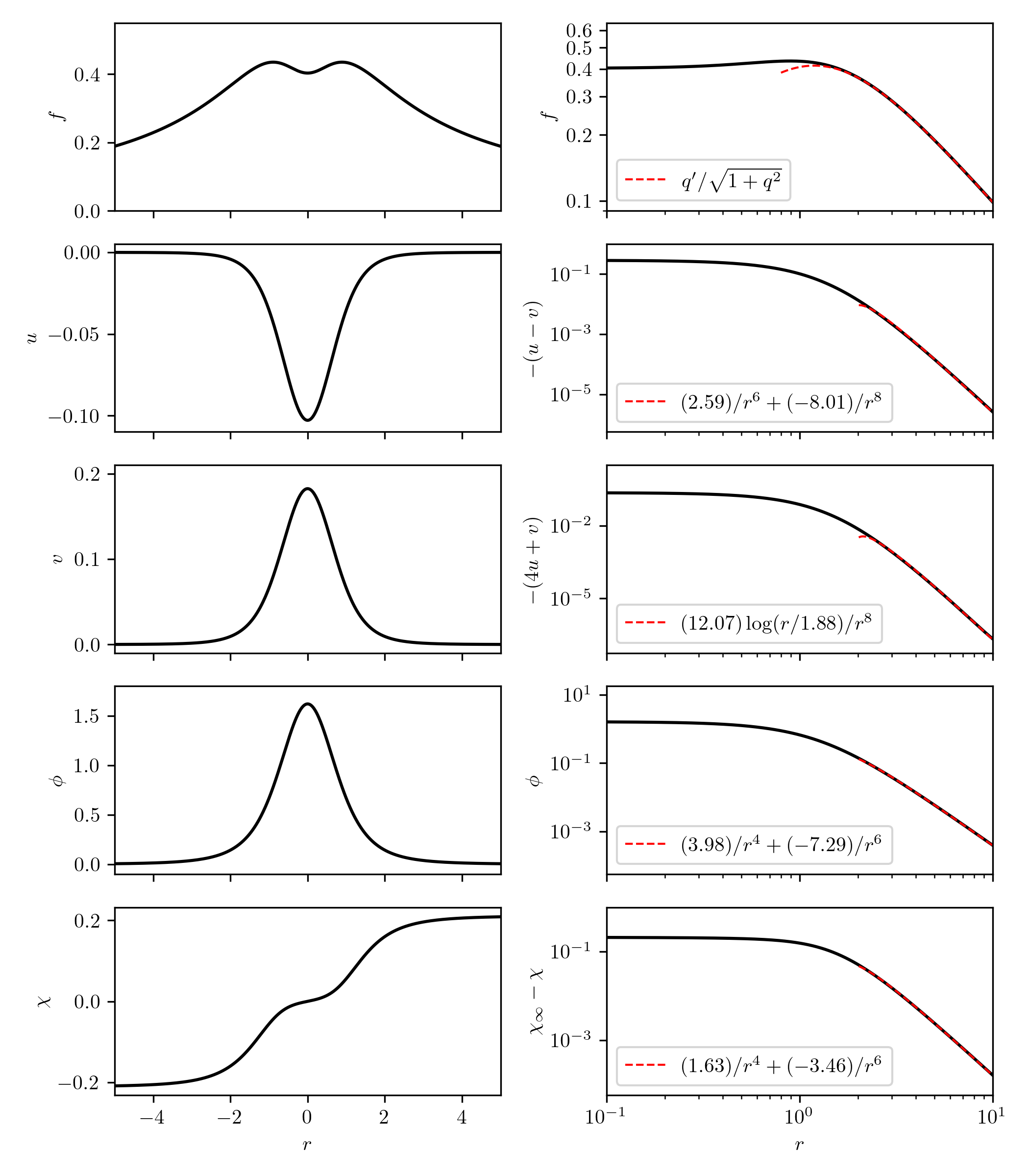}
    \caption{The wormhole solution for $q_0=1$ and $(e^\phi \chi')_0=0.1$. (Left) Field profiles around the wormhole neck. (Right) The profiles match the expected power-law behaviors for $r\to\infty$.}
    \label{fig:T11_profiles}
\end{figure}

In constructing numerical solutions have the freedom to choose coordinates such that
\begin{equation}
    q(r) = \sqrt{q_0^2 + r^2} \,,
\end{equation}
where $q_0$ is the wormhole size (in AdS units) and $r\in\R$ covers the entire geometry. We are interested in AdS solutions for which $f\to 1/q\sim 1/r$ as $r\to\infty$ (recall equation~\eqref{eq:typeIIB_T11_ansatz}). The boundary conditions for the scalars can be found in the following way. By expanding around the minimum of $\mathcal{V}$ and diagonalizing, one finds two mass eigenstates:
\begin{equation}
\begin{aligned}
    u-v &: & \qquad m_1^2 &= 12 \,, & \qquad \Delta_1 &= 6 \,,\\
    4u+v &: & m_2^2 &= 32 \,, & \Delta_2 &= 8 \,.
\end{aligned}
\end{equation}
These are both subject to a source term from the axions in~\eqref{eq:5d_equations} which goes as $\mathfrak{q}_2^2(\partial h)^2\sim \mathfrak{q}_2^2f^{-2}(h')^2\sim \mathfrak{q}_2^2/r^8$ for large $r$. We should thus anticipate $u-v \sim 1/r^6$ and $4u+v \sim \log{r}/r^8$ for $r\to\infty$.\footnote{The $\log{r}$ appears because the axion source terms go as $1/r^8=1/r^{\Delta_2}$ for large $r$, and the resonant response to this source term is not subdominant.} The scalars $\phi,\chi,b,c$ are all massless and will fall off as $1/r^4$.

For fixed $q_0$ we use a shooting method to construct solutions which connect the neck region to the asymptotic AdS region; ``initial'' conditions at $r=0$,
\begin{equation}
\begin{aligned}
    f(0) &= f_0 \,, & \quad u(0) &= u_0 \,, & \quad v(0) &= v_0 \,, & \quad \phi(0) &= \phi_0 \,, & \quad \chi(0) &= 0 \,,\\
    & & u'(0) &= 0 \,, & v'(0) &= 0 \,, & \phi'(0) &= 0 \,, & \chi'(0) &= (\chi')_0 \,,
\end{aligned}
\end{equation}
must be chosen in such a way that, upon integrating the equations of motion out to some $r_\text{max}\gg\max(1,q_0)$, the solutions remain smooth and may be matched onto the AdS boundary conditions just discussed. We can always pick $\phi_0=0$ and use the $\SL(2,\R)$ transformation of equation~\eqref{eq:rescale} to shift $\phi$ and rescale $\mathfrak{q}_2,\chi$ at the end to ensure that $\phi(\infty)=0$ (the combination of initial data $(e^\phi\chi')_0$ is invariant under this rescaling).

Requiring that the solution is smooth at the wormhole neck relates the initial condition for $f$ and the axion charge to the other initial data:
\begin{equation}
\begin{aligned}
    f_0^{-2} &= 3 - \frac{1}{3}q_0^2\mathcal{V}(u_0,v_0) \,,\\
    \mathfrak{q}_2^2 &= e^{-4u_0+\phi_0}q_0^6\left[ 6 + \left(6 - q_0^2(e^\phi\chi')_0^2\right)f_0^{-2}\right] \,.
\end{aligned}
\end{equation}
Of course we require $f_0^{-2}>0$, which mildly restricts the potential choices for $u_0,v_0$. After picking $q_0$ and $(e^\phi\chi')_0$, the values of $u_0,v_0$ are uniquely determined by the requirement that the solution extends smoothly to $r\to\infty$: see Appendix~\ref{app:massiveIIA} for a discussion of how the appropriate values can be found. In this way we are able to construct a two-parameter family of wormhole solutions, controlled by $q_0$ and $(e^\phi\chi')_0$. For fixed $q_0$, the initial condition $(e^\phi\chi')_0$ is 1-to-1 with the boundary value $\chi_\infty$.

The results of this method for $q_0=1$ and $(e^\phi\chi')_0=0.1$ are shown in figure~\ref{fig:T11_profiles}. We see that for $r\to\infty$ the expected power-law fall-offs are respected, allowing for a smooth gluing to the asymptotic AdS solutions. The solutions are qualitatively similar as $q_0$ and $(e^\phi\chi')_0$ are varied: $\chi$ is always monotonic and $(-u)$, $v$ and $\phi$ are always peaked at $r=0$, decreasing monotonically to zero for $r\to\infty$. This allows us to easily determine the field ranges of $u$, $v$ and $\phi$ in terms of the initial data.

\begin{figure}[t]
    \centering
    \includegraphics[width=\textwidth]{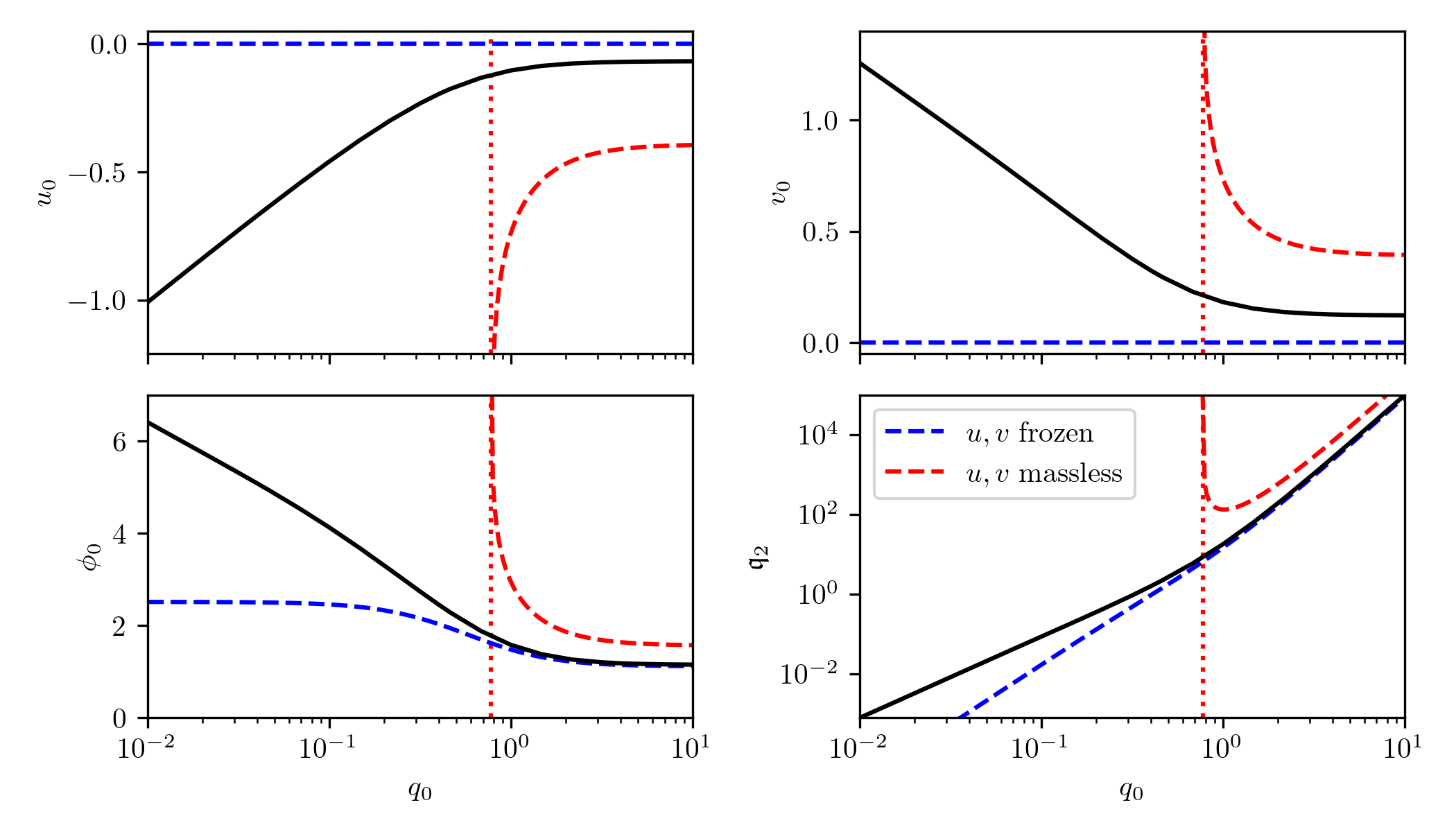}
    \caption{Scalar field ranges and charge as functions of the wormhole size for $\chi=0$. In blue/red are shown the corresponding values for the two ``approximations'' where $u,v$ are frozen/massless. The dotted line indicates the value $q_0=0.7727$, below which the GS ``approximation'' with $u,v$ massless becomes singular: for $q_0\lesssim 0.7727$ all of the scalars traverse larger and larger distances (in Planck units).}
    \label{fig:T11_ranges_chi=0}
\end{figure}

\begin{figure}[t]
    \centering
    \includegraphics[width=\textwidth]{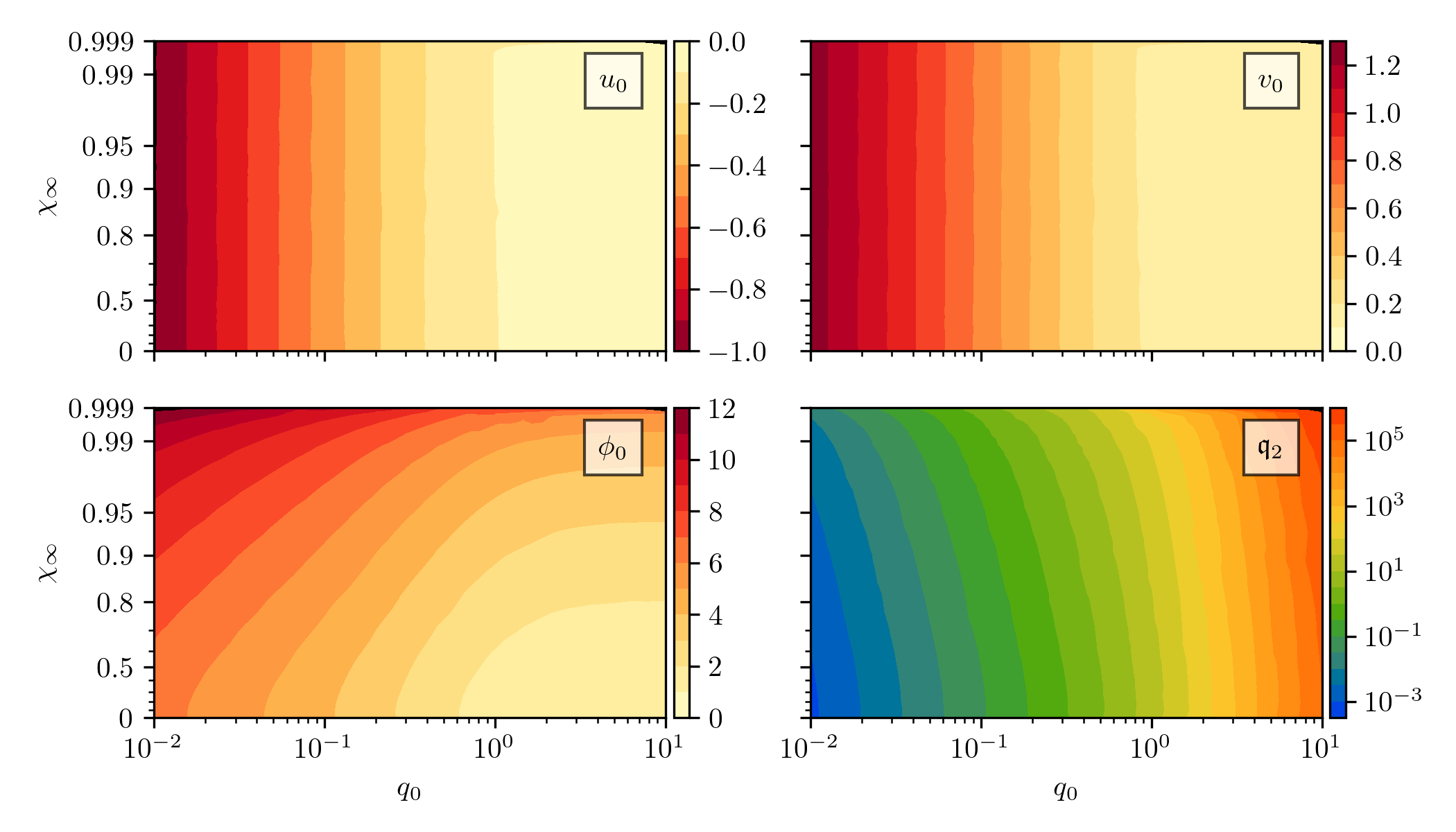}
    \caption{Scalar field ranges and axion charge as functions of the wormhole size, $q_0$, and boundary value of the axion, $\chi_\infty$ (note the non-linear vertical axis). For $q_0\lesssim 1$ or $\chi_\infty\sim 1$ the scalars traverse large (in Planck units) distances in field space.}
    \label{fig:T11_ranges}
\end{figure}

For $(e^\phi\chi')_0=0$ the axion is identically zero, $\chi=0$, and we can make a comparison with the two GS ``approximations'' of section~\ref{sec:GSapprox} where $u,v$ are artificially frozen to zero or made massless. Figure~\ref{fig:T11_ranges_chi=0} shows how the scalar field ranges and axion charge depend on the wormhole size: for large wormholes they are safely sandwiched by the two ``approximations.'' However, since the GS ``approximation'' with $u,v$ massless is regular only for $q_0>0.7727$, the scalars can -- and do -- grow larger and larger as one takes the wormhole size well below the AdS scale, $q_0\ll 1$.
These issues with small wormholes are only exacerbated by taking $\chi\neq0$, which could be anticipated because the ($\phi,\chi$)-only wormholes discussed in section~\ref{sec:GSapprox} are singular. In figure~\ref{fig:T11_ranges} we see how the field ranges and axion charge vary in the two-dimensional parameter space. The profiles for $f$, $u$ and $v$ (which together control the 10d geometry) are minimally affected for $\chi_\infty>0$, but the dilaton field range is increased.
We emphasize that for any fixed $q_0,\chi_\infty$ the solution can be brought under arbitrarily good control with curvatures well below the string scale and weak coupling thoughout by choosing $\ell$ and $g_\text{s}$ in equation~\eqref{eq:typeIIB_T11_ansatz} appropriately. However, for fixed $\ell$ or $g_\text{s}$ only wormholes which are larger than some threshold size are under control.

We can also identify a sharp discontinuity in the limit $q_0\to0$, $\chi_\infty\to1$ which corresponds to (formally) recovering the supersymmetric solution. As we have seen, the massive scalars $u,v$ must be sourced to achieve smooth wormhole solutions, whereas $u$ and $v$ are identically zero when supersymmetry is preserved. Since $u,v$ grow to size $\mathcal{O}(\log{\frac{1}{q_0}})$ (see figure~\ref{fig:T11_ranges_chi=0}) in the neck region $|r|\lesssim q_0\ll 1$, the solutions can converge to $u=v=0$ at best point-wise.

\subsubsection{BPS bound violation}

As described in detail in~\cite{Klebanov:1998hh}, type IIB on $T^{1,1}$ is dual to a 4d, $\N=1$ quiver CFT with two gauge nodes. The scalar field combinations dual to the theta-angles $\theta_i$ and coupling constants $g_i$ have been identified in~\cite{Klebanov:1998hh} and are often written as follows:
\begin{equation}\label{eq:dict}
\begin{aligned}
    \theta_1 + \theta_2 &= 2\pi i\, C_0 = -\frac{2\pi\chi}{g_\text{s}} \,, & \frac{1}{g^2_1} + \frac{1}{g^2_2} &= \frac{1}{4\pi g_\text{s}} \,,\\
    \theta_1 - \theta_2 &= 2\pi i\int_{S^2}C_2=
    -\frac{2\pi c}{g_\text{s}^{1/2}} \,, & \qquad \frac{1}{g^2_1} - \frac{1}{g^2_2} &= \frac{1}{4\pi}\int_{S^2}B_2 =\frac{g_\text{s}^{1/2} b}{4\pi} \,.
\end{aligned}
\end{equation}
The corresponding operators are conventionally denoted as    
\begin{equation}
\begin{aligned}
    \mathcal{O}_\Phi &= \operatorname{Tr}(F_1\wedge{\star F_1} + F_2\wedge{\star F_2}) \,,\\
    \mathcal{O}_{B_2} &= \operatorname{Tr}(F_1\wedge{\star F_1} - F_2\wedge{\star F_2}) \,,\\
    \mathcal{O}_{C_0} &= \operatorname{Tr}(F_1\wedge F_1 + F_2\wedge F_2) \,,\\
    \mathcal{O}_{C_2} &= \operatorname{Tr}(F_1\wedge F_1 - F_2\wedge F_2) \,.
\end{aligned}
\end{equation}
There are bounds which arise from the positivity of the operators,
\begin{equation}
    0\leq \frac{1}{2}\big\langle\!\Tr[(F_i\pm{\star F_i})\wedge\star(F_i\pm{\star F_i})]\big\rangle = \big\langle\!\Tr(F_i\wedge{\star F_i})\big\rangle \pm \big\langle\!\Tr(F_i\wedge F_i)\big\rangle\,,
\end{equation}
for either choice of sign and both $i=1,2$, with equality corresponding to (anti-)instanton solutions with ${\star F_i}=F_i$ (${\star F_i}=-F_i$). 
These positivity bounds can be rewritten in the following way:
\begin{equation}\label{eq:posbounds}
\begin{aligned}
    i=1\;:\qquad \langle\mathcal{O}_\Phi\rangle + \langle\mathcal{O}_{B_2}\rangle &\geq \langle\mathcal{O}_{C_0}\rangle + \langle\mathcal{O}_{C_2}\rangle \,,\\
    i=2\;:\qquad \langle\mathcal{O}_\Phi\rangle - \langle\mathcal{O}_{B_2}\rangle &\geq \langle\mathcal{O}_{C_0}\rangle - \langle\mathcal{O}_{C_2}\rangle \,.
\end{aligned}
\end{equation}
Before we study the corresponding positivity bounds for our numerical solutions we suggest a small change to the commonly-understood holographic dictionary~\eqref{eq:dict}. The issue is the dual of the $C_2$ axion, which was suggested to be the difference in $\theta$-angles of the two gauge groups. This cannot be quite correct since the corresponding operator densities are differences between Pontryagin indices and hence a conserved charge, whereas the corresponding supergravity charge is neither conserved nor gauge-invariant. However the simple change
\begin{equation}
    \int_{S^2} C_2 \quad\longrightarrow\quad \int_{S^2} C_2 - C_0\int_{S^2} B_2
\end{equation}
does the job. Note that in our backgrounds this combination comes from a local two-form $\tilde{C}_2$ defined as $\d{\tilde{C}_2} = \d{C_2}-C_0\,\d{B_2}$ because the right hand side is closed ($F_1\wedge H_3=0$).

\begin{figure}[t]
    \centering
    \includegraphics[width=\textwidth]{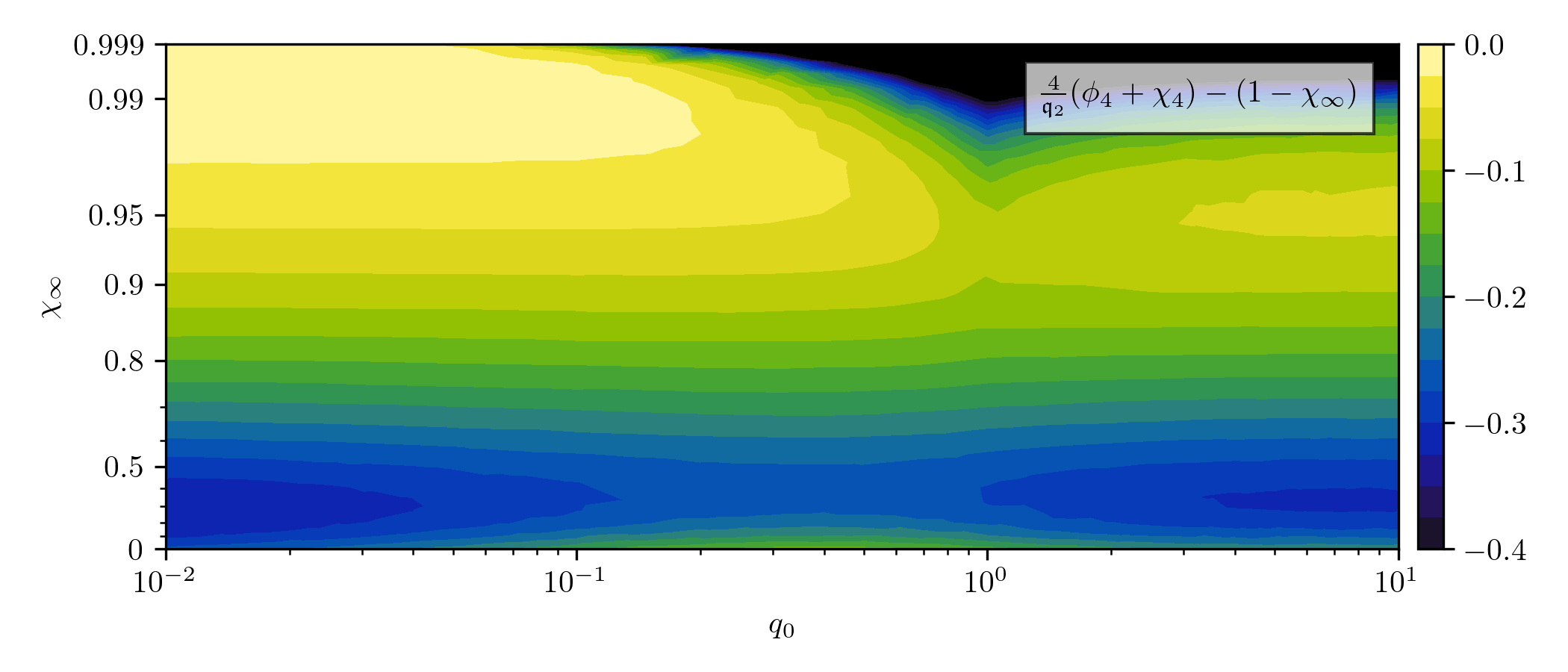}
    \caption{The combination of boundary data from equation~\eqref{eq:BPSrescaled} for the constructed wormholes, which should be positive to satisfy the BPS bound.}
    \label{fig:T11_BPS}
\end{figure}

We now evaluate the positivity bounds, temporarily lifting the boundary condition on the string dilaton. First, unpacking the solutions for $b,c$ in equation~\eqref{eq:axion_charges} in terms of the axion charges gives\footnote{The $1/4$ comes from solving $h'=\frac{f}{q^4}=\frac{1}{r^5}+\cdots$ for $r\to\infty$: $h=h_\infty - \frac{1}{4r^4} + \cdots$.}
\begin{equation}
    b_4 = \frac{1}{4}e^{\phi_\infty}(\mathfrak{q}_1-\chi_\infty\mathfrak{q}_2) \,, \qquad c_4 - \chi_\infty b_4 = \frac{1}{4}e^{-\phi_\infty}\mathfrak{q}_2\,,
\end{equation}
where $(\bullet)_4$ indicates the coefficient of $1/r^4$ in the limit $r\to\infty$. These are exactly the two gauge-invariant combinations of $\mathfrak{q}_1$ and $\mathfrak{q}_2$. Note that the expression on the right is essentially what appears in the expansion of $\tilde{C}_2$ and indeed gives the conserved charge $\mathfrak{q}_2$. We then find
\begin{equation}
\begin{aligned}
    \langle\mathcal{O}_\Phi\rangle &= \phi_4 \,,\\
    \langle\mathcal{O}_{B_2}\rangle &= e^{-\phi_\infty/2}b_4 = \frac{1}{4}e^{\phi_\infty/2}(\mathfrak{q}_1 - \chi_\infty\mathfrak{q}_2) \,,\\
    \langle\mathcal{O}_{C_0}\rangle &= -e^{\phi_\infty}\chi_4 \,,\\
    \langle\mathcal{O}_{\tilde{C}_2}\rangle &= -e^{\phi_\infty/2}(c_4- \chi_\infty b_4) = -\frac{1}{4}e^{-\phi_\infty/2}\mathfrak{q}_2 \,.\\
\end{aligned}
\end{equation}
The bounds of equation~\eqref{eq:posbounds} become
\begin{equation}
    \phi_4 + e^{\phi_\infty}\chi_4 \geq \pm \frac{1}{4}\big[e^{\phi_\infty/2}(\mathfrak{q}_1 - \mathfrak{q}_2\chi_\infty) + e^{-\phi_\infty/2}\mathfrak{q}_2 \big]\,,
\end{equation}
which passes multiple checks. Firstly, both bounds ($\pm$) are saturated iff both sides vanish, which occurs exactly when $\chi = e^{-\phi} + \frac{\mathfrak{q}_1}{\mathfrak{q}_2}$, the supersymmetric solution we saw before. In addition, each side is invariant under both the ``rescaling symmetry'' and all three shift symmetries. Since we choose $\phi_\infty=\mathfrak{q}_1=0$, we just need to check
\begin{equation}
    \phi_4 + \chi_4 \geq \pm \frac{\mathfrak{q}_2}{4}\big( 1 - \chi_\infty \big)
\end{equation}
for the numerical solutions. It helps to rewrite this as
\begin{equation}\label{eq:BPSrescaled}
    \frac{4}{|\mathfrak{q}_2|}(\phi_4 + \chi_4) \pm \big( 1 - \chi_\infty \big) \geq 0
\end{equation}
so that all terms are $\mathcal{O}(1)$; this combination (with lower sign, which is more constraining) is shown in figure~\ref{fig:T11_BPS} for the numerically constructed wormhole solutions. We see that the bound is violated for all $q_0,\chi_\infty$ but approaches saturation (from the ``wrong'' direction) in the limit $q_0\to0$, $\chi_\infty\to1$ in which the supersymmetric solutions are formally recovered. This confirms what has been found in~\cite{Katmadas:2018ksp} for GS wormholes inside $\AdS_5\times S^5/\Z_k$, but now with full control over the 10d solution. 

\begin{figure}[t]
    \centering
    \begin{tikzpicture}
        \draw[ultra thick, red!65!black] (-6,-2) -- (-6,1) -- (-5,2) -- (-5,-1) -- cycle;
        \draw[ultra thick, blue!65!black] (-3,-2) -- (-3,1) -- (-2,2) -- (-2,-1) -- cycle;
        \node at (-5.5, 0) {$\boldsymbol{\times}$};
        \node at (-2.5, 0) {$\boldsymbol{\times}$};

        \draw[dashed, gray] (0,-2.5) -- (0,2.5);
        \node at (0,-3) {};
        
        \draw[thick] (4,0.25) arc (90:270:0.077 and 0.27);
        \draw[dotted] (4,0.25) arc (90:-90:0.077 and 0.27);
        
        \draw[ultra thick, red!65!black] (2,-2) -- (2,1) -- (3,2) -- (3,-1) -- cycle;
        \fill[white] (2.95,-0.43) rectangle (3.05,0.43);
        \draw[ultra thick] plot[smooth] coordinates {(2.5,0.7) (3,0.43) (3.5,0.3) (4,0.27) (4.5,0.3) (5,0.43) (5.5,0.7)};
        \draw[ultra thick] plot[smooth] coordinates {(2.5,-0.7) (3,-0.43) (3.5,-0.3) (4,-0.27) (4.5,-0.3) (5,-0.43) (5.5,-0.7)};
        \draw[ultra thick, blue!65!black, fill=white] (5,-2) -- (5,1) -- (6,2) -- (6,-1) -- cycle;
        \draw[ultra thick] (5.5,-0.7) arc (270:90:0.2 and 0.7);
    \end{tikzpicture}
    \caption{Two-boundary solutions with asymptotic charges of opposite sign (red/blue). Shown are a disconnected instanton/anti-instanton pair (left), each saturating their respective BPS bounds, and a wormhole ``bound state'' of lower action (right).}
    \label{fig:two_bdy}
\end{figure}
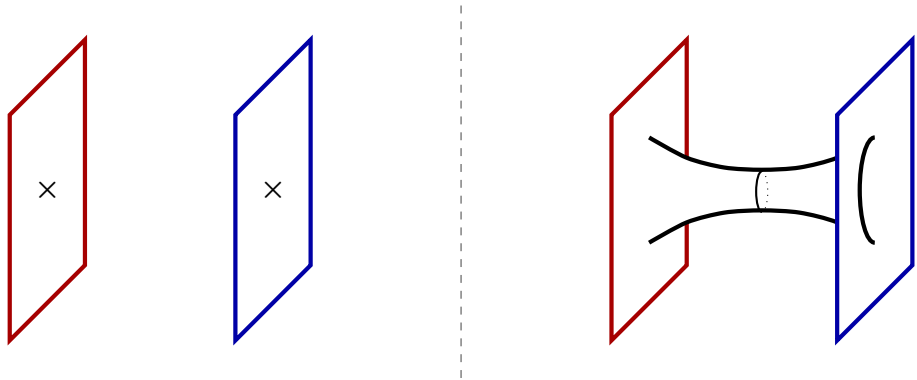

We now speculate on the meaning of this positivity violation.  The wormhole solutions we have constructed have two asymptotic regions, related by parity, $r\to-r$, or equivalently $\chi\to-\chi$ and $\mathfrak{q}_2\to-\mathfrak{q}_2$ (i.e.\ $C_0\to-C_0$ and $C_2\to-C_2$). The BPS bound for the opposite, $r\to-\infty$ region would be
\begin{equation}
    \phi_4 - \chi_4 \geq \pm\frac{\mathfrak{q}_2}{4}(1+\chi_\infty)\,,
\end{equation}
and is saturated by the \emph{other} supersymmetric solutions, namely anti-instantons which have
\begin{equation}
    \chi + e^{-\phi} = \frac{\mathfrak{q}_1}{\mathfrak{q}_2} = 0 \,.
\end{equation}
We can thus understand the $q_0\to0$ limit of the constructed wormhole solutions as pinching off into two \emph{disconnected} spacetimes, being an instanton/anti-instanton pair of opposite charges. In figure~\ref{fig:two_bdy} are sketched both the supersymmetric and wormhole solutions with two boundaries and fixed asymptotic charges. As we saw above, the wormhole configuration violates the BPS bound and has lower action than its supersymmetric counterpart.

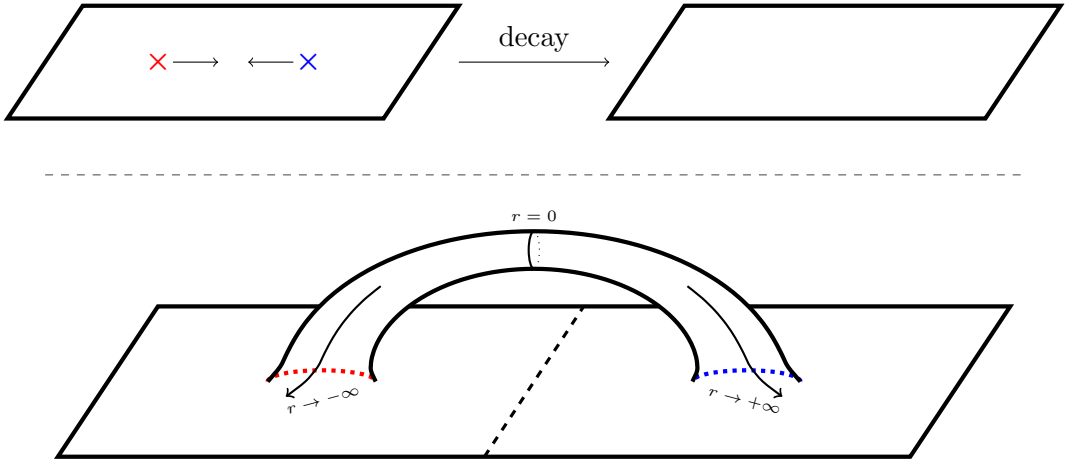
\begin{figure}[t]
    \centering
    \begin{tikzpicture}
        \draw[ultra thick] (-7,3.5) -- (-6,5) -- (-1,5) -- (-2,3.5) -- cycle;
        \node[red] at (-5,4.25) {$\boldsymbol{\times}$};
        \node[blue] at (-3,4.25) {$\boldsymbol{\times}$};
        \draw[->] (-4.8,4.25) -- (-4.2,4.25);
        \draw[->] (-3.2,4.25) -- (-3.8,4.25);

        \draw[ultra thick] (1,3.5) -- (2,5) -- (7,5) -- (6,3.5) -- cycle;
        \draw[->] (-1,4.25) -- node[above] {decay} (1,4.25);

        \draw[dashed, gray] (-6.5,2.75) -- (6.5,2.75);
        \node at (0,-1.75) {};

        \draw[ultra thick] (-6.33,-1) -- (-5,1) -- (6.33,1) -- (5,-1) -- cycle;
        \draw[very thick, dashed] (-0.66,-1) -- (0.66,1);

        \fill[white] (-2.8516,0.95) rectangle (-1.6649,1.05);
        \fill[white] (1.6649,0.95) rectangle (2.8516,1.05);
        
        \draw[ultra thick, dotted, red] (-3.54075,0) arc (180:0:0.718 and 0.15);
        \draw[ultra thick, dotted, blue] (3.54075,0) arc (0:180:0.718 and 0.15);
        
        \draw[thick] (0,2) arc (90:270:0.0714 and 0.25);
        \draw[dotted] (0,2) arc (90:-90:0.0714 and 0.25);
        
        \draw[ultra thick] plot[smooth] coordinates {(2.10463,0.)(2.16825,0.147026)(2.16108,0.292635)(2.12019,0.435427)(2.05355,0.574025)(1.96439,0.707095)(1.85469,0.833355)(1.72609,0.95159)(1.58016,1.06066)(1.4185,1.15952)(1.2428,1.2472)(1.05487,1.32288)(0.856571,1.38582)(0.649874,1.43541)(0.436814,1.47118)(0.21948,1.49278)(0.,1.5)(-0.21948,1.49278)(-0.436814,1.47118)(-0.649874,1.43541)(-0.856571,1.38582)(-1.05487,1.32288)(-1.2428,1.2472)(-1.4185,1.15952)(-1.58016,1.06066)(-1.72609,0.95159)(-1.85469,0.833355)(-1.96439,0.707095)(-2.05355,0.574025)(-2.12019,0.435427)(-2.16108,0.292635)(-2.16825,0.147026)(-2.10463,0.)};
        \draw[ultra thick] plot[smooth] coordinates {(3.54075,0.)(3.37619,0.196034)(3.27893,0.390181)(3.17592,0.580569)(3.05297,0.765367)(2.90611,0.942793)(2.7345,1.11114)(2.53865,1.26879)(2.31977,1.41421)(2.07956,1.54602)(1.82006,1.66294)(1.54359,1.76384)(1.25266,1.84776)(0.949959,1.91388)(0.638323,1.96157)(0.320673,1.99037)(0.,2.)(-0.320673,1.99037)(-0.638323,1.96157)(-0.949959,1.91388)(-1.25266,1.84776)(-1.54359,1.76384)(-1.82006,1.66294)(-2.07956,1.54602)(-2.31977,1.41421)(-2.53865,1.26879)(-2.7345,1.11114)(-2.90611,0.942793)(-3.05297,0.765367)(-3.17592,0.580569)(-3.27893,0.390181)(-3.37619,0.196034)(-3.54075,0.)};

        \node[above] at (0,2) {\tiny$r=0$};
        \draw[thick, ->] plot[smooth] coordinates {(2.03865,1.26879)(2.2345,1.11114)(2.40611,0.942793)(2.55297,0.765367)(2.67592,0.580569)(2.77893,0.390181)(2.87619,0.196034)(3.04075,0.)(3.3,-0.2)};
        \draw[thick, ->] plot[smooth] coordinates {(-2.03865,1.26879)(-2.2345,1.11114)(-2.40611,0.942793)(-2.55297,0.765367)(-2.67592,0.580569)(-2.77893,0.390181)(-2.87619,0.196034)(-3.04075,0.)(-3.3,-0.2)};
        \node[rotate=-15] at (2.8,-0.25) {\tiny$r\to+\infty$};
        \node[rotate=15] at (-2.8,-0.25) {\tiny$r\to-\infty$};
    \end{tikzpicture}
    \caption{Single-boundary solutions with vanishing asymptotic charges, to be contrasted with figure~\ref{fig:two_bdy}. (Top) An instanton/anti-instanton pair of opposite charges will attract and annihilate to form empty AdS$_5$. (Bottom) The wormhole solution formed by gluing the $r\to\infty$ and $r\to-\infty$ regions is a ``dipole'' source for $\chi,c$ and satisfies the dual positivity bounds.}
    \label{fig:one_bdy}
\end{figure}

If we instead consider gluing the two asymptotic regions of the wormhole solutions together, then the state of affairs is very different: see figure~\ref{fig:one_bdy}. The analogous instanton/anti-instanton solution with one boundary is unstable and decays to empty AdS. In addition, we find that the single-boundary wormhole solutions now \emph{do} satisfy the dual positivity bounds. This is ultimately due to the fact that the wormhole acts as a dipole source for the RR fields $C_0$ and $C_2$ ($\chi$ and $c$) so that the corresponding one-point functions vanish. Said differently, since these fields are odd functions of $r$ in the two-boundary solutions, upon gluing only the odd multipole moments survive. The opposite is true for the NSNS fields, and so we can write
\begin{equation}
\begin{aligned}
    \langle\mathcal{O}_\Phi\rangle_\text{glued} &= \phi_4 \,,\\
    \langle\mathcal{O}_{B_2}\rangle_\text{glued} &= e^{\phi_\infty/2}b_4 \,,\\
    \langle\mathcal{O}_{C_0}\rangle_\text{glued} &= 0 \,,\\
    \langle\mathcal{O}_{\tilde{C}_2}\rangle_\text{glued} &= 0 \,.
\end{aligned}
\end{equation}
The $C_0,C_2$ axions help to support the wormhole solutions without contributing to the positivity bounds,
\begin{equation}
    \phi_4 \geq \pm \frac{\mathfrak{q}_2}{4}\chi_\infty \,,
\end{equation}
which are now satisfied by a wide margin for all of the constructed wormholes. However, note that this gluing procedure requires the solutions be altered, as one expects a small attraction between the two wormhole mouths. But, similar to the dilute instanton gas approximation, we ignore this and consider these as approximate solutions.


\section{Conclusion}
\label{sec:conclusion}

In this paper we have described, for the first time, Euclidean wormholes of Giddings-Strominger type from a 10d viewpoint. This has furthermore lead to new Euclidean axion wormhole solutions in AdS compactifications from considering type IIB supergravity on $T^{1,1}$. Let us summarize our findings.
\begin{itemize}
    \item Both the Euclidean wormhole solutions in flat space and AdS can be tuned parametrically into the supergravity regime such that they can be trusted as \emph{bona fide} string backgrounds. This is despite the scalars flowing through distances of a few Planck lengths.
    \item Since the extremal limits of the wormholes give BPS Euclidean branes there is a na\"ive paradox how wormholes can be regarded in this 10d picture since branes do not have regular over-extremal limits. We noticed that supergravity evades this paradox by only allowing regular wormholes when enough branes intersect in such a way that a Lorentzian interpretation in terms of over-extremal branes is impossible.
    \item We have found entirely new and explicit, yet numerically constructed, wormhole solutions in AdS compactifications. In the main text we discussed the solutions in type IIB on $\AdS_5\times T^{1,1}$ which implies a concrete holographic dual in the form of the Klebanov-Witten gauge theory. In an appendix we present a somewhat simpler solution in massive type IIA, but since the vacuum is non-SUSY a holographic dual is not known. An important ingredient in both constructions is the inclusion of massive scalars.
    \item Since the wormholes are sourced by scalars whose dual operators are known, we could verify that the wormholes violate operator positivity in the dual CFT. This was also found earlier in~\cite{Katmadas:2018ksp} but for wormholes inside $\AdS_5\times S^5/\Z_k$ where the orbifolding obscures a simple 10d understanding.
    \item We have then speculated how changing these wormhole solutions into wormholes connected to the same boundary would escape violating the positivity bound. This suggests, once again, that AdS/CFT with multiple boundaries is not well understood. 
\end{itemize}
These findings have also raised further puzzles. Our wormholes traversed super-Planckian distances in field space, but not parametrically large, so there is no sharp contradiction with the distance conjecture. Yet we have made clear that, strictly speaking, there is no clear definition of the distance conjecture in Euclidean spacetime signature since the scalar manifold then becomes of indefinite signature. Should only the distance in dilaton space be measured? A more striking puzzle is the understanding of the two-boundary wormholes we constructed. The violation of operator positivity shows that the holographic dictionary in the presence of two boundaries is not fully understood. It is our hope that our explicit wormholes can help further studies in this regard.

Finally, we note that the explicit AdS axion wormholes that we have constructed have little bearing on discussions of baby universes, $\alpha$-parameters and factorization, other than reinforcing that the issues they raise are alive and well in string theory. It has been argued in~\cite{McNamara:2020uza} that the baby universe Hilbert space should be one-dimensional, meaning that there is a great redundancy amongst states of different topologies. Indeed, in~\cite{Marolf:2020xie} it was argued that a dramatic reduction in $\dim\mathcal{H}_\text{BU}$ compared to na\"ive expectations can occur due to a large number of null states. There, this effect was made explicit in toy models where the sum over topologies could be done exactly. Importantly, it was shown that this effect is completely invisible in perturbation theory, e.g.\ by truncating a topological expansion or working only at low-energies, as we have done here. Also, it is perhaps suggestive that for our solutions the limit of small wormholes in AdS units (but still very large in string or Planck units) where one would hope to integrate out their effects \emph{\`a la} Coleman is \emph{exactly} where the scalars $u,v,\phi$ controlling the 10d curvatures and couplings become large, invalidating the low-energy effective description.


\section*{Acknowledgments}

We thank both Pablo Soler and Jake McNamara for useful discussions.
The work of GJL is supported in part by MEXT Leading Initiative for Excellent Young Researchers Grant Number JPMXS0320210099.
The work of GS is supported in part by the DOE grant DE-SC0017647.
The work of TVR is supported by the C16/16/005 grant of the KULeuven.


\appendix

\section{Type II equations of motion}
\label{app:typeII_eoms}

Here we recall the usual 10d Einstein frame equations of motion for bosonic modes in type II supergravity. For type IIB we have
\begin{equation}
\begin{aligned}
    \d{\star\d{\Phi}} &= -\tfrac{1}{2}e^{-\Phi}H_3\wedge{\star H_3} + \sum_{p=1,3}\tfrac{5-p}{4}\,e^{(5-p)\Phi/2}F_p\wedge{\star F_p} \,,\\
    \d{(e^{-\Phi}{\star H_3})} &= F_3\wedge F_5 + e^\Phi F_1\wedge{\star F_3} \,,\\
    \d{(e^{2\Phi}{\star F_1})} &= -e^\Phi H_3\wedge{\star F_3} \,,\\
    \d{(e^\Phi{\star F_3})} &= -H_3\wedge F_5 \,,\\
    \d{\star F_5} &= H_3\wedge F_3 \,,\\
    2R_{MN} &= \partial_M\Phi\partial_N\Phi + e^{-\Phi}\left(\tfrac{1}{2}(H_3)_{MN}^2 - \tfrac{1}{24}g_{MN}H_3^2\right)\\
    &\qquad + \sum_{p=1,3}e^{(5-p)\Phi/2}\left(\tfrac{1}{(p-1)!}(F_p)_{MN}^2 - \tfrac{p-1}{8\cdot p!}g_{MN}F_p^2\right) + \tfrac{1}{2\cdot 4!}(F_5)_{MN}^2 \,,
\end{aligned}    
\end{equation}
where
\begin{equation}
    (F_p)_{MN}^2 = (F_p)_{MP\cdots Q}(F_p)_N{}^{P\cdots Q} \,, \qquad F_p^2 = (F_p)_{M\cdots Q}(F_p)^{M\cdots Q} \,,
\end{equation}
and similarly for $H_3$. The Bianchi identities are
\begin{equation}
    \d{H_3} = 0 \,, \qquad \d{F_1} = 0 \,, \qquad \d{F_3} = H_3\wedge F_1 \,, \qquad \d{F_5} = H_3\wedge F_3 \,,
\end{equation}
and self-duality of $F_5$ is imposed as a constraint by hand.
For massive type IIA the equations of motion read
\begin{equation}
\begin{aligned}
    \d{\star\d{\Phi}} &= -\tfrac{1}{2}e^{-\Phi}H_3\wedge{\star H_3} + \sum_{p=0,2,4}\tfrac{5-p}{4}\,e^{(5-p)\Phi/2}F_p\wedge{\star F_p} \,,\\
    \d{(e^{-\Phi}{\star H_3})} &= \frac{1}{2}F_4\wedge F_4 + e^{3\Phi/2}F_0\wedge{\star F_2} + e^{\Phi/2}F_2\wedge{\star F_4} \,,\\
    \d{(e^{3\Phi/2}{\star F_2})} &= -e^{\Phi/2}H_3\wedge{\star F_4} \,,\\
    \d{(e^{\Phi/2}{\star F_4})} &= -H_3\wedge F_4 \,,\\
    2R_{MN} &= \partial_M\Phi\partial_N\Phi + e^{-\Phi}\left(\tfrac{1}{2}(H_3)_{MN}^2 - \tfrac{1}{24}g_{MN}H_3^2 \right)\\
    &\qquad + \sum_{p=0,2,4}e^{(5-p)\Phi/2}\left(\tfrac{1}{(p-1)!}(F_p)_{MN}^2 - \tfrac{p-1}{8\cdot p!}g_{MN}F_p^2\right) \,,
\end{aligned}
\end{equation}
(where $(F_0)_{MN}^2=0$ is understood) and are accompanied by the following Bianchi identities:
\begin{equation}
    \d{H_3} = 0 \,, \qquad \d{F_0} = 0 \,, \qquad \d{F_2} = H_3\wedge F_0 \,, \qquad \d{F_4} = H_3\wedge F_2 \,.
\end{equation}
In appendix~\ref{app:massiveIIA} we will take $F_2=H_3=0$, in which case the above reduce to
\begin{equation}\label{eq:IIA_eoms}
\begin{aligned}
    \d{\star\d{\Phi}} &= \sum_{p=0,4}\tfrac{5-p}{4}\,e^{(5-p)\Phi/2}F_p\wedge{\star F_p} \,,\\
    0 &= F_4\wedge F_4 \,,\\
    \d{(e^{\Phi/2}{\star F_4})} &= 0 \,,\\
    2R_{MN} &= \partial_M\Phi\partial_N\Phi + \sum_{p=0,4}e^{(5-p)\Phi/2}\left(\tfrac{1}{(p-1)!}(F_p)_{MN}^2 - \tfrac{p-1}{8\cdot p!}g_{MN}F_p^2\right) \,,
\end{aligned}
\end{equation}
and $\d{F_0}=\d{F_4}=0$.


\section{Massive type IIA on \texorpdfstring{$S^3\times S^3$}{S3xS3}}
\label{app:massiveIIA}

We saw in section~\ref{sec:flatspace_WHs} that type IIA on a 6-torus allows for regular, 4d flat space wormholes. In this section we discuss an ``AdS analogue'' of these solutions by considering massive type IIA on $S^3\times S^3$. Here we describe a one-parameter family of AdS wormholes for which the scalars remain bounded and we have parametric control over the 10d solutions. The construction parallels that of the type IIB wormholes of section~\ref{sec:AdS_WHs}, but with fewer scalar fields and the wormhole size as the only adjustable parameter we will use these solutions to exemplify the numerics and shooting method used in constructing both type IIB and massive type IIA wormholes.

As in section~\ref{sec:flatspace_WHs}, we will immediately set $H_3=F_2=0$. There is a \emph{non-supersymmetric} $\mathrm{AdS}_4\times S^3\times S^3$ solution of~\eqref{eq:IIA_eoms} supported by both $F_0$- and $F_4$-flux:
\begin{equation}\label{eq:AdS4soln}
\begin{aligned}
    \d{s_{10}^2} &= \ell^2\left(\frac{\d{r^2}}{1+r^2} + r^2\,\d{\Omega_3^2}\right) + \ell^2\big(\d{\Omega_{3,1}^2} + \d{\Omega_{3,2}^2}\big) \,,\\
    e^\Phi &= g_\text{s} \,,\\
    F_0 &= \sqrt{2}\,\ell^{-1}g_\text{s}^{-5/4} \,,\\
    F_4 &= i\sqrt{10}\,\ell^{-1}g_\text{s}^{-1/4}\,\left(\ell^4\vol_4\right) \,.
\end{aligned}
\end{equation}
This solution was first found in~\cite{Romans:1985tz} (see also~\cite{Koerber:2010rn}). 
The ansatz we use for wormholes inside this background is the following,
\begin{equation}\label{eq:typeIIA_S3xS3_ansatz}
\begin{aligned}
    \d{s_{10}^2} &= \ell^2e^{-3u}\left[f(r)^2\,\d{r^2} + q(r)^2\,\d{\Omega_3^2}\right] + \ell^2e^u\big(\d{\Omega_{3,1}^2} + \d{\Omega_{3,2}^2}\big) \,,\\
    e^\Phi &= g_\text{s}e^\phi \,,\\
    F_0 &= \sqrt{2}\,\ell^{-1}g_\text{s}^{-5/4} \,,\\
    F_4 &= i\sqrt{10}\,\ell^{-1}g_\text{s}^{-1/4}e^{-9u-\phi/2}\,\left(\ell^4\vol_4\right) + \frac{i}{\sqrt{2}}\,\ell^3g_\text{s}^{-1/4}\,\d{\chi}\wedge\big(\vol_{3,1} + \vol_{3,2}\big) \,,
\end{aligned}
\end{equation}
with $u,\phi\to0$ for $r\to\infty$. In this ansatz we already restrict to an isotropic axion for simplicity, analogous to what was done at the end of section~\ref{sec:IIA_reduction}. It is straightforward to extend the analysis that follows to two axions, one for each $S^3$, and two radions.

The 4d equations of motion for $g_{\mu\nu}$, $u$, $\phi$ and $\chi$ which result from the above ansatz can be derived from the 4d action
\begin{equation}\label{eq:4Daction}
    S_4 = \frac{1}{2\kappa_4^2}\int\Big({\star \mathcal{R}} - 6\,\d{u}\wedge{\star\d{u}} - \frac{1}{2}\d{\phi}\wedge{\star\d{\phi}} + \frac{1}{2}e^{-3u+\phi/2}\d{\chi}\wedge{\star\d{\chi}} - {\star\mathcal{V}}\Big)
\end{equation}
with scalar potential
\begin{equation}
    \mathcal{V} = -12e^{-4u} + 5e^{-9u-\phi/2} + e^{-3u+5\phi/2} \,.
\end{equation}
Note that we scaled out $\ell$ and $g_\text{s}$; by design, the scalar potential takes on its minimum value of $\mathcal{V}=-6$ at $u=\phi=0$. Upon finding solutions in terms of these rescaled fields, one can always arrange for weak coupling and small curvature everywhere by choosing $\ell,g_\text{s}$ in~\eqref{eq:typeIIA_S3xS3_ansatz} appropriately.

Because of the scalar potential in the action of equation~\eqref{eq:4Daction}, there are no GS wormhole solutions. We can, however, gain some insight into whether regular wormholes exist by artificially making $u$ and $\phi$ massless via the \emph{ad hoc} substitution $\mathcal{V}\to-6$ just like what was done in section~\ref{sec:GSapprox}. Having done this, the 4d action can be rewritten as
\begin{equation}\label{eq:S3xS3_massless_action}
    \frac{1}{2\kappa_4^2}\int\Big({\star(\mathcal{R}+6)} - \frac{1}{2}\delta_{ij}\d{\varphi^i}\wedge{\star\d{\varphi^j}} + \frac{1}{2}e^{\varphi^1}\d{\chi}\wedge{\star\d{\chi}}\Big) \,,
\end{equation}
where
\begin{equation}
    \varphi^1 = -3u+\frac{\phi}{2} \,, \qquad \varphi^2 = \sqrt{3}\left(u + \frac{\phi}{2}\right) \,.
\end{equation}
Thus in this ``approximation'' there is a GS wormhole solution with $\varphi^2$ decoupled and $\varphi^1,\chi$ comprising an axio-dilaton pair with $\beta=1$, which is regular everywhere for any choice of wormhole size $q_0$ (equivalently, any $\c<0$):
\begin{equation}
    \frac{1}{\beta^2} > \frac{3}{4} = \left(\frac{D_4(0)}{2\pi}\right)^2 \geq \left(\frac{D_4(q_0)}{2\pi}\right)^2 \,, \qquad \c = -12q_0^4(1+q_0^2) \,.
\end{equation}
In restoring the correct functional form for $\mathcal{V}$ it is clear that regular wormhole solutions will continue to exist, since the potential only helps to further stabilize $u$ and $\phi$ near their asymptotic values of zero.

\subsection{Numerical solutions}

\begin{figure}[t]
    \centering
    \includegraphics[width=\textwidth]{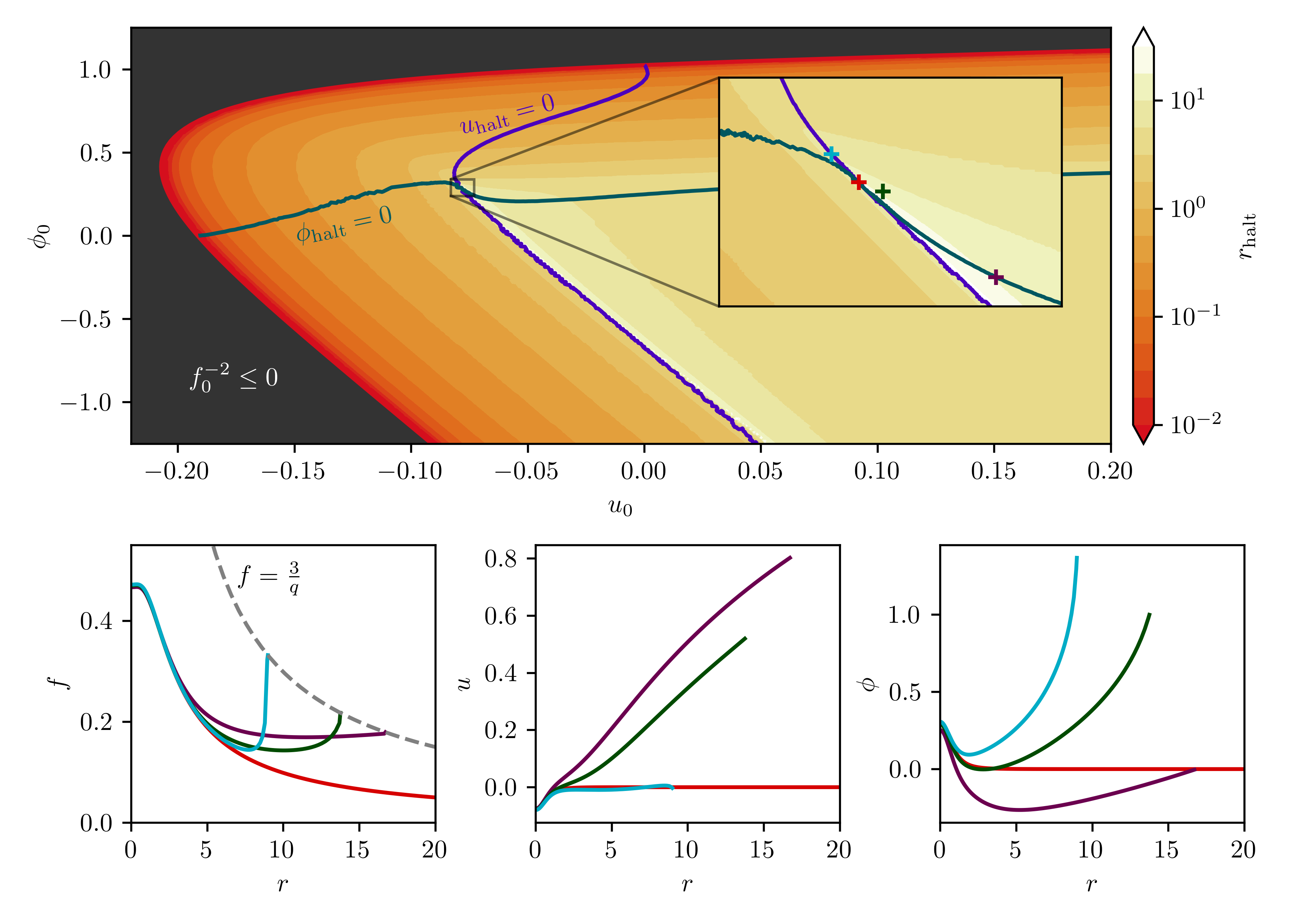}
    \caption{(Top) Shooting method parameter space for $S^3\times S^3$ solutions with $q_0=1$ and $r_\text{max}=50$, showing how $r_\text{halt}$ depends on $u_0$ and $\phi_0$. Green and purple contours indicate parameters for which $\phi_\text{halt}$ and $u_\text{halt}$ are zero, respectively. The optimal initial conditions are given by the red `$+$' in the inset, for which $r_\text{halt}=r_\text{max}$ and $u,\phi\to0$. (Bottom) Profiles for $f$, $u$ and $\phi$ corresponding to all four choices of parameters indicated by `$+$'s in the inset.}
    \label{fig:S3xS3_paramspace}
\end{figure}

As in section~\ref{sec:wh_typeIIB} we again choose
\begin{equation}
    q(r) = \sqrt{q_0^2 + r^2}
\end{equation}
and solve the axion equation of motion in terms of a charge $\mathfrak{q}$ as
\begin{equation}
    {\star(e^{-3u+\phi/2}\,\d{\chi})} \equiv \mathfrak{q}\,\vol_3 \qquad\Longleftrightarrow\qquad \d{\chi} = e^{3u-\phi/2}\mathfrak{q}\,\d{h}\,,
\end{equation}
where here the harmonic function $h(r)$ solves $h'=f/q^3$, i.e.\ $\star\d{h} = \vol_3$. This leaves only $f$, $u$ and $\phi$ to be found numerically. Here the string dilaton has mass of order the AdS scale, ultimately originating from the non-zero background fluxes. Expanding the potential around $u=\phi=0$, one identifies the two mass eigenstates:
\begin{equation}
\begin{aligned}
    6u-5\phi &: &\qquad m_1^2 &= 6 \,, &\qquad \Delta_1 &= \tfrac{3}{2} + \sqrt{\tfrac{9}{4}+6} = 4.3722\ldots \,,\\
    10u+\phi &: & m_2^2 &= 20 \,, & \Delta_2 &= \tfrac{3}{2} + \sqrt{\tfrac{9}{4}+20} = 6.2169\ldots \,.
\end{aligned}
\end{equation}
Both $u$ and $\phi$ are sourced by the axion with a term that goes as $\mathfrak{q}^2(\partial\chi)^2\sim\mathfrak{q}^2f^{-2}(h')^2\sim\mathfrak{q}^2/r^6$ for large $r$. Therefore we should find that for $r\to\infty$ the scalars go to zero as
\begin{equation}\label{eq:S3xS3_powerlaws}
    6u-5\phi = \frac{A_1}{r^{\Delta_1}} + \mathfrak{q}^2\,\frac{A_2}{r^6} + \cdots \,, \qquad
    10u+\phi = \mathfrak{q}^2\,\frac{A_3}{r^6} + \frac{A_4}{r^{\Delta_2}} + \cdots \,,
\end{equation}
with $A_i$ some numerical coefficients.

The shooting method used is structured as follows. For fixed $q_0>0$ and $r_\text{max}\gg\max(1,q_0)$ the equations of motion are solved for $r\leq r_\text{max}$ with initial conditions
\begin{equation}
    f(0) = f_0 \,, \qquad u(0)=u_0 \,, \qquad \phi(0)=\phi_0 \,, \qquad f'(0)=u'(0)=\phi'(0)=0 \,.
\end{equation}
Constructing a perturbative solution around $r=0$ shows that $f_0$ and $\mathfrak{q}$ are determined in terms of $u_0,\phi_0$:
\begin{equation}\label{eq:S3xS3_f0q}
\begin{aligned}
    f_0^{-2} &= 2 - \frac{q_0^2}{2}\mathcal{V}(u_0,\phi_0) \,,\\
    \mathfrak{q}^2 &= 4e^{-3u_0+\phi_0/2}q_0^4\left(1 + f_0^{-2}\right) \,.
\end{aligned}
\end{equation}
Generic choices for $u_0,\phi_0$ lead to singular geometries where $f(r)$ diverges: the integration is halted at $r=r_\text{halt}$ if $f(r_\text{halt})>\frac{3}{q}$. The value of $r_\text{halt}\leq r_\text{max}$ in the $(u_0,\phi_0)$ parameter space, along with curves where the final values of the fields, ${u_\text{halt}=u(r_\text{halt})}$ and ${\phi_\text{halt}=\phi(r_\text{halt})}$, vanish are shown in figure~\ref{fig:S3xS3_paramspace} for $q_0=1$.

One may engineer an objective function $\mathcal{U}$ which depends implicitly on $u_0,\phi_0$ through $r_\text{halt},u_\text{halt},\phi_\text{halt}$ so that minimizing $\mathcal{U}$ ensures the solution is smooth and may be glued onto the AdS boundary conditions. We choose
\begin{equation}
    \mathcal{U}(u_0,\phi_0) = \begin{cases}
        A_1 + |f_0^{-2}| & f_0^{-2} \leq 0\\
        1 + A_2\big(\frac{r_\text{max}}{r_\text{halt}}-1\big) + u_\text{halt}^2 + \phi_\text{halt}^2 & r_\text{halt} < r_\text{max}\\
        1 - \frac{1}{1 + [(\phi-1.2u)_\infty^\text{est}]^2 + [(u+0.1\phi)_\infty^\text{est}]^2} & r_\text{halt} = r_\text{max}
    \end{cases}
\end{equation}
with $A_1,A_2$ two large, fixed constants. The values $(\phi-1.2u)_\infty^\text{est}$ and $(u+0.1\phi)_\infty^\text{est}$ are estimates for the boundary values of the mass eigenstates, leveraging the known power-law behavior in equation~\eqref{eq:S3xS3_powerlaws}:
\begin{equation}
\begin{aligned}
    (\phi-1.2u)_\infty^\text{est} &= (\phi-1.2u)_\text{halt} + \frac{r_\text{halt}}{\Delta_1}(\phi'-1.2u')_\text{halt} \,,\\
    (u+0.1\phi)_\infty^\text{est} &= (u+0.1\phi)_\text{halt} + \frac{r_\text{halt}}{6}(u'+0.1\phi')_\text{halt} \,.
\end{aligned}
\end{equation}
The objective function has a minimal value of zero where $r_\text{halt}=r_\text{max}$ and both extrapolated values $(\phi-1.2u)_\infty^\text{est}$ and $(u+0.1\phi)_\infty^\text{est}$ are zero. Standard optimization techniques, such as gradient descent or the Nelder--Mead method, applied to $\mathcal{U}$ yield the optimal initial conditions. By successively increasing $r_\text{max}$ and extrapolating to $r_\text{max}=\infty$, one may quickly determine $u_0,\phi_0$ very accurately; for example, the initial conditions found for $q_0=1$ are
\begin{equation}
\begin{aligned}
       u_0 &= \mathtt{-0.07891\;87232\ldots} \;,\\
    \phi_0 &= \mathtt{+0.29432\;16217\ldots} \;,
\end{aligned}
\end{equation}
corresponding to the red `$+$' and red profiles in figure~\ref{fig:S3xS3_paramspace}.

\begin{figure}[p]
    \centering
    \includegraphics[width=\textwidth]{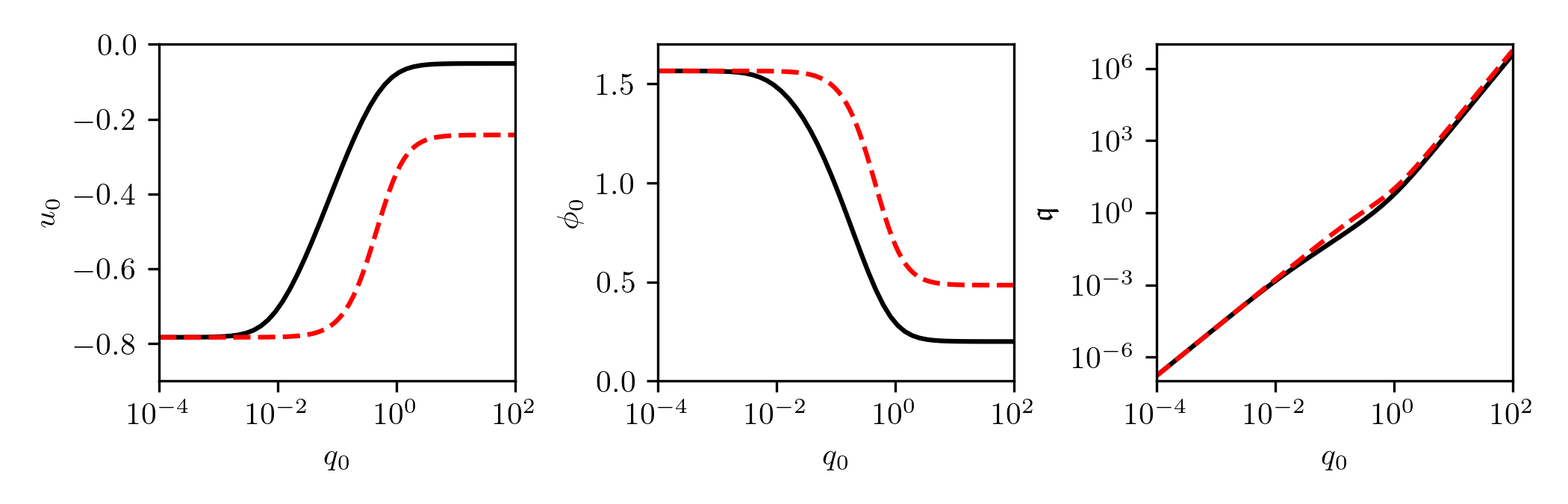}
    \caption{Scalar field ranges and axion charge as a function of the wormhole size. In red are shown the field ranges for the GS solution corresponding to the action of equation~\eqref{eq:S3xS3_massless_action} where $u,\phi$ are artificially made massless. As expected, the field ranges are smaller than those of the ``approximation'' since the potential helps to stabilize $u,\phi$ to zero for $r\to\infty$.}
    \label{fig:S3S3_ranges}
\end{figure}

We have thus found a one-parameter family of solutions, controlled by $q_0$, for which the scalar field ranges and axion charge are easily determined in terms of the optimized initial conditions. Figure~\ref{fig:S3S3_ranges} shows how these characteristics of the wormhole depend on the wormhole size; for all $q_0$ the field ranges are bounded by that of the GS ``approximation'' corresponding to~\eqref{eq:S3xS3_massless_action}, as expected. In particular, even for wormholes much smaller than the AdS scale the scalars always lie in the bounded regions $u\in[{-\frac{1}{2}}\log\sec(\frac{\sqrt{3}\pi}{4}),0]$ and $\phi\in[0,\log\sec(\frac{\sqrt{3}\pi}{4})]$. This is in stark contrast to the solutions for $T^{1,1}$ of section~\ref{sec:wh_typeIIB} for which the scalars become larger and larger for ever-smaller wormholes.

\begin{figure}[p]
    \centering
    \includegraphics[width=\textwidth]{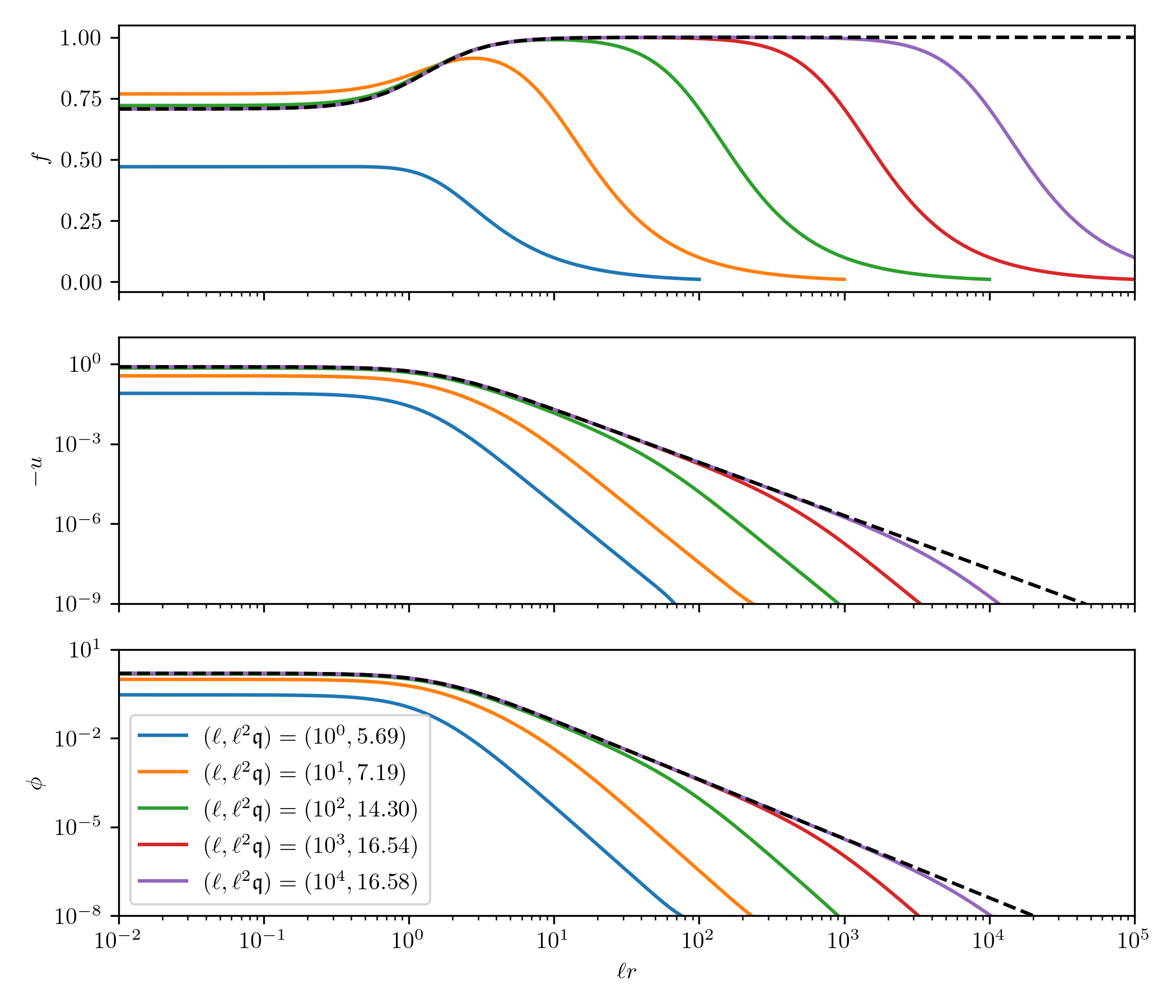}
    \caption{Profiles for $a_0=\ell q_0=1$ fixed and $\ell\in\{1,10,10^2,10^3,10^4\}$, showing point-wise convergence to a flat space, GS wormhole solution ($\beta=1$) shown in black. The growing intermediate region $a_0\lesssim\rho\lesssim\ell$ where $f\approx 1$ becomes the asymptotically-flat boundary region in the strict $\ell\to\infty$ limit.}
    \label{fig:S3S3_flat}
\end{figure}

\subsection{Relationship with flat space solutions}

Because the scalars $u$ and $\phi$ are well-behaved in the limit $q_0\to0$ where the wormhole is made small in AdS units, we can make the flat space limit of these solutions precise. Returning to equation~\eqref{eq:typeIIA_S3xS3_ansatz} and introducing the rescaled variables $\rho$ and $a$ through $r=\rho/\ell$ and $q=a/\ell$ and using stereographic projection to write the 3-sphere metrics as ${\ell^2\,\d{\Omega_{3,i}^2} = \big(1+\frac{|{\bf x}|^2}{4\ell^2}\big)^{-2}(\d{\bf x}\cdot\d{\bf x})}$,
we can then take $\ell\to\infty$ to find
\begin{equation}
\begin{aligned}
    \d{s_{10}^2} &= e^{-3u}\left(f^2\,\d{\rho^2} + a^2\,\d{\Omega_3^2}\right) + e^u\,\d{s_{\R^6}^2} \,,\\
    e^\Phi &= g_\text{s}e^\phi \,,\\
    F_0 &= 0 \,,\\
    F_4 &= 0 + \frac{i}{\sqrt{2}}\,g_\text{s}^{-1/4}\,\d{\chi}\wedge\big(\d{x^1}\wedge\d{x^2}\wedge\d{x^3} + \d{x^4}\wedge\d{x^5}\wedge\d{x^6}\big) \,,
\end{aligned}
\end{equation}
In particular the background $F_0,F_4$ fluxes vanish and one has ``zoomed in'' on the north poles of the $S^3$s to obtain an $\R^6$ factor. After quotienting to obtain a 6-torus, we arrive at \emph{non-supersymmetric} solutions of GS type, similar to those in section~\ref{sec:flatspace_WHs}.
We can explicitly see this flat space limit in the numerical solutions by looking at a family of solutions with fixed $a_0=\ell q_0$ and increasing $\ell$. In figure~\ref{fig:S3S3_flat} we see several rescaled profiles for $a_0=1$ which exhibit the point-wise convergence to a flat space GS solution with $\beta=1$. An intermediate region, $a_0\lesssim \rho\lesssim \ell$, emerges in which $f\approx 1$ and $u,\phi\sim \rho^{-2}$, before matching on to the AdS solution for $\rho\gtrsim\ell$. In the strict $\ell\to\infty$ limit this intermediate region becomes the asymptotically-flat region at infinity.


\bibliographystyle{utphys}
\bibliography{refs}

\end{document}